\documentclass[11pt]{JHEP3}

\usepackage{amsmath,amssymb,graphicx,epsfig,subeqnarray,psfrag,wrapfig}

\newcommand{\be}{\begin{equation} }
\newcommand{\ee}{\end{equation} }
\newcommand{\ba}{\begin{eqnarray}}
\newcommand{\ea}{\end{eqnarray}}

\def\E{{\cal E}}

\def\p{\partial}
\def\s{\sigma}
\def\d{\delta}
\def\nn{\nonumber}

\def\I_M{{I_{\scriptscriptstyle M\times M}}}

\preprint{KIAS-P07076}

\title{Higgs Structures of Dyonic Instantons}

\author{Min-Young Choi$^{*a}$, Kyung Kiu Kim$^{\dag b}$, Choonkyu Lee$^{*c}$,
and Ki-Myeong Lee$^{\dag d}$   \\
\\
$^*$ Department of Physics and Astronomy and Center for
Theoretical
Physics,\\ Seoul National University, Seoul 151-747, Korea \\
$^\dag$ Korea Institute for Advanced Study, Seoul, 130-722,
Korea\\ $~$ \\ $^a$ E-mail: witt1010@snu.ac.kr \\ $^b$ E-mail:
kimeagle@kias.re.kr \\ $^c$ E-mail: cklee@phya.snu.ac.kr \\ $^d$
E-mail: klee@kias.re.kr}

\abstract{We study Higgs field configurations of dyonic instantons
in spontaneously broken $(4+1)$-dimensional Yang-Mills theory. The
adjoint scalar field solutions to the covariant Laplace equation
in the ADHM instanton background are constructed in general
noncanonical basis, and they are used to study explicitly the
Higgs field configurations of dyonic instantons when the gauge
fields are taken by Jackiw-Nohl-Rebbi instanton solutions. For
these solutions corresponding to small instanton number we then
consider in some detail the zero locus of the Higgs field, which
describes the cross section of supertubes connecting parallel
D4-branes in string theory. Also the information on the Higgs
zeroes is used to discuss the residual gauge freedom concerning
the Jackiw-Nohl-Rebbi solutions.}

\begin{document}

\section{Introduction}

Instanton solutions of 4-dimensional Euclidean Yang-Mills theory
are also known to play a role as solitons in certain spontaneously
broken $(4+1)$-dimensional gauge theory. Corresponding objects,
first discussed by Lambert and Tong \cite{tong}, are called dyonic
instantons, for stable instantons in the broken vacuum must come
with nonzero electric charge. In type IIA string theory the
D-brane interpretation of dyonic instantons can be found in a
supertube \cite{mt,mnt,kmpw} which connects parallel D4-branes
lying close to each other. See also
Refs.~\cite{bl,zam,seok,leeyee,hash} for discussions relevant to
dyonic instantons from the latter perspective.

Classical dyonic instanton solutions, as described by Yang-Mills
gauge field $A_\mu$ and Higgs scalar $\phi$ (which are functions
of four spatial coordinates $x_\mu$), satisfy Bogomol'nyi-type
equations: especially, $A_\mu(x)$ satisfy the usual self-duality
conditions appropriate to Yang-Mills instantons
\be\label{sd} F_{\mu\nu}= {^*}F_{\mu\nu} \ee
while the scalar fields (in the adjoint representation) $\phi(x)$
obey the covariant Laplace equation in the background of the
instanton
\be\label{cL} D_\mu D_\mu \phi = 0~.\ee
The gauge fields can thus be described by the ADHM construction
\cite{adhm,cws,cori}, which includes the Jackiw-Nohl-Rebbi (JNR)
instanton solutions \cite{jnr} as (explicit) special cases. The
adjoint Higgs solutions to (\ref{cL}) are easy to find when the
gauge fields are taken by 't Hooft solutions \cite{thooft}, but,
with the JNR instanton backgrounds, complete expressions are known
only for the two-instanton case \cite{seok}. The latter
expressions were found using the construction of
Refs.~\cite{dkm,kms,dhkm}, where the appropriate form for the
Higgs field was identified when the ADHM data for the background
gauge fields was presented in the canonical basis. It is much
desirable to have more systematic understanding on the solutions
to (\ref{cL}) in non-'t Hooft-type instanton backgrounds since the
zero locus of these Higgs fields is known to have direct
connection with the cross section of the (noncollapsed)
supertubes.

In the present work we have two goals. The first is to provide a
fuller analysis on the solutions of (\ref{cL}), in their general
structure and also by obtaining new explicit solutions in JNR
backgrounds. The second is to use such knowledge on the Higgs
configurations, especially their zero locus, to clarify further
various issues related to the supertube interpretation of dyonic
instantons. In this work we find Higgs fields satisfying the
covariant Laplace equation in general ADHM backgrounds (given in
arbitrary basis, canonical or not), directly from the asymptotic
behavior of the known scalar propagator \cite{cws,cgt} in ADHM
backgrounds. It is seen that the complex structure Higgs solutions
have is entirely due to the same matrix factor that also enters
the propagator expression for adjoint scalar. To find explicit
Higgs solutions in JNR backgrounds, we can now utilize our
construction directly (without going through the awkward procedure
\cite{seok} as needed to express JNR instanton solutions using
canonical ADHM data first); fully explicit forms of the Higgs
solutions can be produced this way. Then, as in Ref.~\cite{seok},
the zero locus of the Higgs fields can be studied. On the special
significance carried by these Higgs zeroes, some explanations will
be offered below.

As noted in Ref.~\cite{seok}, the zero locus of the Higgs field is
the magnetic monopole string along which the unbroken $U(1)$
magnetic flux emerges. A loop-like boundary of a tubular D2-brane
connecting two parallel but separated D4-branes would appear as
the magnetic string on D4 worldvolume. Let us recall that a
supertube is made of a tube of D2-brane with fundamental strings
(F1) lying along the tube direction and D0-branes spread along the
D2-brane such that the F1 string number times the D0-brane number
density remains constant \cite{mnt}. Also the straight D2-brane
connecting two parallel but separated D4-branes appears as a
$U(1)$ magnetic monopole string of $(4+1)$-dimensional Yang-Mills
theory. Thus the zero locus is the direct indicator of the way a
supertube connects two D4-branes, and on this aspect we shall
expand the discussions of Ref.~\cite{seok} further by studying the
zeroes of the explicit Higgs solutions obtained in this work. A
supertube with large D0 and F1 charges can have arbitrary cross
section. Similarly, the number of moduli parameters for the
magnetic monopole string would increase with the instanton number.
If $\kappa$ is the instanton number, then the dimension of
instanton moduli space is $8\kappa$. In the Coulomb phase, the
fixed electric charge gives one constraint and the corresponding
coordinate is cyclic. Thus dyonic instantons of the instanton
number $\kappa$ in the center of mass frame has $8\kappa-6$
independent parameters which also characterize the shape of the
magnetic monopole string.

Explicit instanton solutions/Higgs configurations are hard to
obtain in general so that the structure of the zero locus as the
representation of the cross section of supertubes is difficult to
get in full detail. For the JNR-type dyonic instanton with the
instanton number $\kappa$, the number of relevant parameters is
given by $5\kappa+6$ or $5\kappa+7$, depending on whether the JNR
position parameters are aligned along a circle (or a straight
line) or not. While the number of parameters in the JNR-type
solutions is less than $8\kappa$ (as appropriate for the most
general $SU(2)$ dyonic instanton solutions), the Higgs field of
the JNR-type solutions have an intricate structure to warrant
further investigations. In this work, we find explicit Higgs
solutions for the JNR-type dyonic instanton with the instanton
numbers $\kappa=3$ and $\kappa=4$ and obtain explicit expressions
for the electric charge with some small value of $\kappa$. We then
elaborate on the structures of the zero locus of the thus-obtained
Higgs field, expanding the similar analysis done in
Ref.~\cite{seok} significantly. The global gauge orientation of
the instanton configurations with respect to the asymptotic Higgs
expectation value plays an important role in changing the shape of
the zero locus. Also analyzed is the reduction of the number of
moduli which happens when all the JNR position parameters are
aligned on a circle \cite{jnr}, by utilizing the gauge invariance
of the zero locus of the Higgs field.

This paper is organized as follows. In Sec.~2 we identify the
Higgs field forms of dyonic instantons, that go with ADHM
instanton fields in general basis. (We here assume spontaneously
broken $SU(2)$ gauge theory). The electric charge is also computed
for general dyonic instantons. Section 3 is devoted to finding the
explicit Higgs solutions in JNR backgrounds, and we see some
regular patterns emerging. Based on the findings of Sec.~3, we
then analyze in Sec.~4 the Higgs zero locus structure in some
detail and examine the related D4-brane-supertube configurations.
We also use the information on the Higgs zeroes to study the
residual gauge freedom that enters the general JNR instanton
solutions. Section 5 contains conclusion and discussions for
future study. There are two appendices. In Appendix A we provide a
direct verification for our Higgs field construction in general
ADHM backgrounds. Appendix B contains certain technical parts
relevant in the computation of electric charge.

\section{The ADHM Construction of Dyonic Instantons in General
Basis}

We will start with presenting the Bogomol'nyi-type equations
describing dyonic instantons in $(4+1)$-dimensional
Yang-Mills-Higgs theory \cite{tong}. We here have the energy
functional
\be {\cal E} = -\frac1{e^2}\int d^4x~{\rm tr}\left\{E_\mu^2
+\frac12 F_{\mu\nu}^2  + (D_t \phi)^2 + (D_\mu
\phi)^2\right\}~,\ee
where $\mu=0,1,2,3$ is a spatial index, $D_\mu\equiv
\p_\mu+[A_\mu,~]$ ($D_t$ likewise), $E_\mu \equiv F_{\mu t}= D_\mu
A_t - \p_t A_\mu$, and $\phi$ denote adjoint scalars. Both
$(A_t,A_\mu)$ and $\phi$ are taken to be antihermitian matrices in
the space of gauge group generators. Then, completing the square
in the usual fashion and using the Gauss law
\be\label{gauss} D_\mu E_\mu -[\phi, D_t \phi]=0~,\ee
it is possible to rearrange this energy functional into the form
\be\label{energy} \E=-\frac1{e^2}\int d^4x~{\rm tr}\left\{(E_\mu
-D_\mu \phi)^2 +\frac14 (F_{\mu\nu}-{^*}F_{\mu\nu})^2+(D_t
\phi)^2\right\}+\frac{8\pi^2}{e^2}\kappa +\frac{2\rm v}{e^2}
Q_e~,\ee
where the quantities
\be \kappa = -\frac1{16\pi^2}\int d^4x~{\rm
tr}(F_{\mu\nu}{^*}F_{\mu\nu}) \ee
and
\be\label{charge} Q_e = -\int d^4x ~\p_\mu \frac1{\rm v}{\rm
tr}(\phi E_\mu) \ee
denote the instanton number and the electric charge, respectively.
In (\ref{charge}) v denotes a constant Higgs vacuum expectation
value, with ${\rm v}^2 = \lim_{|x|\rightarrow\infty} (-2\:{\rm
tr}\:\phi^2)$. When the values of $\kappa$($\geq 0$) and
$Q_e$($\geq 0$) are given, we obtain the energy bound $\E \geq
\left(\frac{8\pi^2}{e^2}\kappa +\frac{2\rm v}{e^2} Q_e\right)$
from (\ref{energy}). Further, this energy bound is saturated by
the static field configurations satisfying
\be\label{BPS} F_{\mu\nu} = {^*}F_{\mu\nu}~,~~~E_\mu = D_\mu
\phi~,~~~D_t \phi =0~,\ee
which are BPS equations for dyonic instantons. We can choose a
gauge where $A_t=\phi$ in which case the fields are static in
time.

{} From the first equation of (\ref{BPS}), the (spatial) gauge fields
of dyonic instantons are taken by usual Yang-Mills instanton
solutions. On the other hand, since the electric field $E_\mu$ of
the field configurations should satisfy the Gauss law constraint
(\ref{gauss}), the last two equations of (\ref{BPS}) can be
combined with the Gauss law to give
\be\label{cL2} D_\mu D_\mu \phi =0~.\ee
Hence the adjoint scalar fields of dyonic instantons should
satisfy the covariant Laplace equation in the background of
instanton solutions (and, at spatial infinity, approach the
suitably chosen vacuum expectation values). Any nontrivial
solution to (\ref{cL2}) should give rise to a nonvanishing
electric charge contribution to the energy, i.e., $Q_e\neq 0$.
Authors of Refs.~\cite{tong,seok} obtained explicit dyonic
instanton solutions for some special cases.

The general solution to Yang-Mills self-duality equations
$F_{\mu\nu} = {^*}F_{\mu\nu}$ is given by the ADHM construction
\cite{adhm,cws,cori}. Assuming $SU(2)$ gauge group the ADHM form
for gauge fields $A_\mu (x) \equiv A^a_\mu (x) \frac{\s_a}{2i}$
($\s_a$, $a=1,2,3$ denote three $2\times 2$ Pauli matrices)
corresponding to $\kappa$ instantons is
\be\label{gauge} A_\mu (x) = \sum_{l=0}^\kappa \bar v_l(x)\p_\mu
v_l (x) \equiv v^\dag (x)\p_\mu v(x)~,\ee
where $v_l(x)\equiv v^\mu_l(x) e_\mu$, $\bar v_l(x)\equiv
v^\mu_l(x) \bar e_\mu$ with $e_\mu \equiv (e_0={1}, \vec e=
-i\vec\s)$ and $\bar e_\mu \equiv ({1},i\vec\s)$ (i.e., $v_l(x)$
and $\bar v_l (x)$ are quaternionic objects), and $v(x)$ a
quaternionic $(\kappa+1)$-column vector. Then, because of the
self-duality equations, the quaternionic column vector $v(x)$ is
constrained by the equations
\ba & v^\dag(x)v(x) = 1~,\nonumber\\\label{constraint} &
\Delta^\dag(x) v(x) = v^\dag(x) \Delta (x)= 0~,\ea
where $\Delta$, a quaternionic $(\kappa +1)\times\kappa$ matrix
with linear $x$-dependence
\be\label{Delta} \Delta = B-Cx~,~~~(x\equiv x^\mu e_\mu) \ee
should obey the so-called ADHM constraint: the $\kappa\times\kappa$
matrix $\Delta^\dag(x)\Delta(x)$ must be real (i.e., belong to the
identity element of quaternion) and invertible. The nontrivial part
of this method is to find two constant quaternionic
$(\kappa+1)\times\kappa$ matrices $B$ and $C$, and for $\kappa >3$
the corresponding expressions in their full generality are still not
known.

Now our task is to determine the broken-vacuum solution
$\phi(x)\equiv \phi^a(x)\frac{\s_a}{2i}$ to (\ref{cL2}) with the
above ADHM background for $A_\mu$. By the broken vacuum we mean
that when $A_\mu(x)$ at spatial infinity is described by a pure
gauge of the form
\be\label{puregauge} A_\mu \rightarrow \bar g(\hat x)\p_\mu g(\hat
x)~~,~~~{\rm{as}}~ |x|\rightarrow\infty \ee
(here $\hat x^\mu \equiv x^\mu/|x|$ and $g(\hat x)$ is an
arbitrary unit quaternion), the asymptotic behavior of our Higgs
solution is
\be\label{hvev} |x|\rightarrow \infty ~:~~ \phi(x)\rightarrow
\bar g(\hat x) \phi_0 g(\hat x) \ee
with a {\it constant} matrix $\phi_0$ ($|\phi_0|\equiv
{\sqrt{\phi_0^a \phi_0^a}}={\rm v}$); $\phi_0$ may be identified
with the scalar vacuum expectation value in the gauge where
$A_\mu(x)<{\cal O}(1/|x|)$ as $|x|\rightarrow\infty$. For
$\phi(x)$ belonging to the fundamental representation of $SU(2)$,
this task of solving the covariant Laplace equation would have
been a easy one; the solution with a very simple structure (see
below) was obtained already in Ref.~\cite{godd}. But, in our case,
i.e., with an adjoint scalar, the solution is not so simple due to
certain complications known to occur when fields in higher
dimensional representations are involved. A systematic way to deal
with the problem is the tensor product method of Corrigan et al
\cite{cgt}. This method was used by them to understand the
structure of adjoint scalar propagators \cite{cklee,cws} and also
to obtain the (adjoint representation) solution to massless Dirac
equation in the ADHM background.

Actually, for the solution of (\ref{cL2}), it is unnecessary to go
through the tensor product formalism of Ref.~\cite{cgt} --- we can
instead exploit the known expression for the scalar propagator in
the ADHM instanton background. For the sake of comparison, the
solution to (\ref{cL2}) for both fundamental and adjoint scalar
$\phi$ will be considered. For the fundamental scalar, the
propagator or the inverse covariant Laplacian is \cite{cws,cori}
\be\label{pr12} \Delta^{\left(\frac12\right)}(x,y) =
\frac{v^\dag(x)v(y)}{4\pi^2 (x-y)^2}~, \ee
using $2\times 2$ matrix notations. On the other hand, the
corresponding expression for the adjoint scalar has more
complicated structure \cite{cws,cgt}
\ba \Delta^{(1)}_{ab}(x,y) &=& \frac1{8\pi^2 (x-y)^2}~{\rm tr}\{\s_a v^\dag (x) v(y) \s_b v^\dag (y)v(x)\}\nn \\
&~&+\frac1{8\pi^2}\sum_{ijmn=1}^\kappa {\rm tr}\{C^\dag v(x) \s_a
v^\dag (x)C \}_{ij} f_{ij,mn} {\rm tr}\{C^\dag v(y) \s_b v^\dag
(y)C \}_{mn}~,\nonumber\\\label{propa} ~ \ea
where `${\rm tr}$' refers to the trace of $2\times 2$ matrix
representing quaternionic quantities, and the constants
$f_{ij,mn}$ are specified through the matrix equation
\ba\label{f} f_{ij,mn}L_{mn,rs} &=& \d_{ir}\d_{js} -
\d_{jr}\d_{is}~,
\\\label{L}
 L_{mn,rs} &=& \frac12{\rm tr}\left\{2(C^\dag B)_{mr} (B^\dag C)_{sn}- (C^\dag
C)_{mr} (B^\dag B)_{sn} - (B^\dag B)_{mr} (C^\dag
C)_{sn}\right\}\nonumber \\ &~& -(m \leftrightarrow n)~.\ea
(It was to understand the necessity of the second term on the
right of (\ref{propa}), as first noticed in Ref.~\cite{cklee},
that the tensor product method was originally developed
\cite{cgt}). Now, taking the asymptotic limit $|y|\rightarrow
\infty$ with these propagators, we may express the results by
\begin{subeqnarray}
|y|\rightarrow \infty ~~:~~ \Delta^{\left(\frac12\right)}(x,y) &=&
\frac{\Phi^{\left(\frac12\right)}(x,\hat y)}{|y|^2} +{\cal
O}\left(\frac1{|y|^3}\right)~,\label{scal12}\\
\Delta^{\left(1\right)}_{ab}(x,y) &=&
\frac{\Phi^{\left(1\right)}_{ab}(x,\hat y)}{|y|^2} +{\cal
O}\left(\frac1{|y|^3}\right)~.\label{scal1}
\end{subeqnarray}
Then, from the defining equations satisfied by the propagators,
the functions $\Phi^{\left(\frac12\right)}(x,\hat y)$ and
$\Phi^{\left(1\right)}_{ab}(x,\hat y)$ will have to satisfy the
appropriate covariant Laplace equations, i.e.,
\begin{subeqnarray} D_\mu^{(x)}D_\mu^{(x)} \Phi^{\left(\frac12\right)}(x,\hat y)
&=& 0~,\\ (D_\mu^{(x)}D_\mu^{(x)})_{ab}
\Phi^{\left(1\right)}_{bc}(x,\hat y) &=& 0~,\end{subeqnarray}
the unit vector $\hat y$ serving only as free parameters. Based on
this, we can identify the solution to (\ref{cL2}) (up to a
multiplicative constant) with
\begin{subeqnarray}
\phi^{\left(\frac12\right)}_\alpha (x) &=&
\Phi^{\left(\frac12\right)}_{\alpha\beta}(x,\hat
y)u_\beta~,~~({\rm fundamental}) \\ \phi^{\left(1\right)}_a (x)
&=& \Phi^{\left(1\right)}_{ab} (x,\hat y) u_b~,~~({\rm adjoint})
\end{subeqnarray}
where $u_\beta$ ($\beta=1,2$), $u_b$ ($b=1,2,3$) denote arbitrary
constant isospinor and isovector, respectively. This is our key
observation.

To find Higgs solutions by the above method, we need the
asymptotic behavior of a quaternionic $(\kappa +1)$-column vector
$v(y)$. Because of the first condition  in (\ref{constraint}), we
may here write
\be\label{asym2} v(y) = h_0(\hat y) + h_1(\hat y)\frac1{|y|}+{\cal
O}\left(\frac1{|y|^2}\right)~,~~~{\rm as}~|y|\rightarrow
\infty~.\ee
Then, from (\ref{gauge}), (\ref{puregauge}) and the two conditions
in (\ref{constraint}), we are led to following conclusions:
\begin{itemize}
\item[(i)] $h_0(\hat y)$ can be expressed as
\be\label{asym1} h_0(\hat y)=Vg(\hat y)~~({\rm or}~h_0(\hat y)_l=
V_l\: g(\hat y)~,~l=0,1,\cdots,\kappa)\ee
with a constant quaternionic $(\kappa +1)$-column vector $V$
satisfying the conditions
\be\label{asym4} V^\dag V = 1~,~~C^\dag V=0~.\ee
\item[(ii)] $h_1(\hat y)$ should satisfy the condition $C^\dag
h_1(\hat y)= \hat y B^\dag h_0(\hat y)$, or using (\ref{asym1}),
\be\label{asym3} C^\dag h_1(\hat y) = \hat y B^\dag V g(\hat
y)~.\ee
\end{itemize}
We then notice from (\ref{pr12}) and (\ref{scal12}a) that
$\Phi^{\left(\frac12\right)}(x,\hat y)\propto v^\dag (x) h_0(\hat
y)=v^\dag(x)Vg(\hat y)$ and therefore the fundamental scalar
solution of (\ref{cL2}) is simply
\be\label{12higgs} \phi^{\left(\frac12\right)} (x) = v^\dag(x)
Vg(\hat y)u = v^\dag V\phi_0~,\ee
where we replaced $g(\hat y)u$ by the appropriate vacuum
expectation value $\phi_0$ (in accordance with the asymptotic
requirement $\phi(x)\rightarrow \bar g(\hat x)\phi_0$ as
$|x|\rightarrow \infty$). Our expression (\ref{12higgs}) agrees
with the result obtained in Ref.~\cite{godd}.

For the function $\Phi^{(1)}_{ab}(x,\hat y)$ appropriate to an
adjoint scalar, it is not difficult to see that the first piece in
the corresponding scalar propagator (\ref{propa}) contributes a
term proportional to ${\rm tr}\{\s_a v^\dag(x)Vg(\hat y)\s_b \bar
g(\hat y)V^\dag v(x)\}$. The contribution from the second piece
can also be found easily if one uses the observation (resulting
from (\ref{asym2})--(\ref{asym3}))
\be C^\dag v(y) =\frac1{|y|}\hat y B^\dag Vg(\hat y)+{\cal
O}\left(\frac1{|y|^2}\right)~,~~{\rm as}~|y|\rightarrow\infty \ee
and its consequence
\be {\rm tr}\{C^\dag v(y)\s_b v^\dag (y)C\}_{mn}=\frac1{|y|^2}{\rm
tr}\{B^\dag Vg(\hat y)\s_b \bar g(\hat y)V^\dag B\}_{mn}+{\cal
O}\left(\frac1{|y|^3}\right)~.\ee
(Note that $\hat y\hat{\bar y}=1$). In this way, for the solution
of (\ref{cL2}), we obtain the expression
\ba \phi(x) &\equiv& \phi^{(1)}_a(x) \frac{\s_a}{2i} ~=~
\frac{\s_a}{2i} \Phi^{(1)}_{ab}(x,\hat y) u_b \nn \\
&=& \frac12\sum_a \s_a {\rm tr}\{\s_a v^\dag(x)V\phi_0 V^\dag
v(x)\}+\sum_{ij=1}^\kappa \sum_a \s_a {\rm tr}\{C^\dag v(x)\s_a
v^\dag(x) C\}_{ij}{\cal A}_{ij}~,\nonumber\\ \label{main}~\ea
where we have made the identification (see below)
\be\label{vev} g(\hat y)\vec u \cdot \frac{\vec\s}{2i} \bar g(\hat
y) = \phi_0~,\ee
and introduced the antisymmetric matrix ${\bf\cal A}=({\cal
A}_{ij})$ determined by solving the inhomogeneous linear
simultaneous equations
\be\label{A} L_{mn,rs} {\cal A}_{rs} = {\rm tr}(B^\dag V \phi_0
V^\dag B)_{mn}~.\ee
The identification (\ref{vev}) is the result of comparing the
asymptotic behavior of our expression against (\ref{hvev}). Here,
the second term in (\ref{main}) being ${\cal O}(1/|x|^2)$ as
$|x|\rightarrow\infty$, it suffices to consider the first term
which indeed reduce to $\frac12\sum_a \s_a {\rm tr}\{\s_a \bar
g(\hat x) V^\dag V\phi_0 V^\dag V g(\hat x)\} = \bar g(\hat
x)\phi_0 g(\hat x)$ as $|x|\rightarrow\infty$. We further note
that, since $\frac12\sum_a \s_a {\rm tr}\{\s_a X\}=X$ if $X$ is
traceless, (\ref{main}) can be simplified into the form
\be\label{main2} \phi(x)=v^\dag(x) V\phi_0 V^\dag v(x) -
\sum_{ij=1}^\kappa \left[(v^\dag(x)C)_i (C^\dag v(x))_j - (v^\dag
(x)C)_j (C^\dag v(x))_i\right] {\cal A}_{ij}~.\ee
This is the expression we have been after. We also verified
explicitly that our expression (\ref{main2}) solves (\ref{cL2});
this direct check is not quite trivial (like many other
calculations involving instanton solutions) and so, for interested
readers, we provide some essential steps needed in the
verification in Appendix A.

With the {\it canonical} ADHM data assumed, i.e.,when the
quaternionic $(\kappa +1)\times \kappa$ matrix $\Delta$ is
presented in the form
\be\label{canondata1} \Delta = \left(\begin{array}{c} \Lambda^\mu e_\mu \\ \hline \\
\Omega^\mu e_\mu \\ ~ \end{array}\right)-
\left(\begin{array}{c} 0 \\ \hline \\
I_{\kappa\times\kappa} \\ ~ \end{array}\right)x \equiv B-Cx\ee
($\Lambda^\mu = (\Lambda^\mu_1\cdots \Lambda^\mu_\kappa)$ is a
$\kappa$-row vector, and $\Omega^\mu = (\Omega^\mu_{mn})$ a
$\kappa\times\kappa$ hermitian matrix), (\ref{main2}) reduces to
the result of Refs.~\cite{dkm,kms,dhkm}. This can be seen as
follows. In this canonical basis, the above quaternionic
$(\kappa+1)$ column vector $V$ becomes simply
\be V=  \begin{pmatrix} 1 \\ 0 \\\vdots \\0 \end{pmatrix}~,\ee
and as a result our expression (\ref{main2}) can be organized into
\be\label{canon} \phi(x) = \bar v_0(x)\phi_0 v_0(x)
-2\sum_{ij=1}^\kappa \bar v_i(x) {\cal A}_{ij} v_j(x) = v^\dag (x)
\left(\begin{array}{c|c} \phi_0 & 0 \\ \hline \\ 0 & -2{\cal A} \\
~ & ~ \end{array}\right) v(x)~.\ee
At the same time, the linear equations (\ref{A}) for the matrix
${\cal A}$ (with $L_{mn,rs}$ given in (\ref{L})) can be simplified
using following results valid in this basis:
\ba &~& C^\dag B = \Omega (=\Omega^\mu e_\mu)~,~~B^\dag C =
\Omega^\dag (=\Omega^\mu \bar e_\mu)~,~~B^\dag B= \Lambda^\dag
\Lambda + \Omega^\dag \Omega~,\nonumber\\\label{canondata2} &~&
C^\dag C = I~,~~ B^\dag V = \Lambda^\dag~,~~V^\dag B =
\Lambda~,\ea
and, for the inhomogeneous term in (\ref{A}),
\be\label{inhom} {\rm tr}(B^\dag V\phi_0 V^\dag
B)_{mn}=-2i\eta_{\mu\nu a}\phi_0^a [(\Lambda^\dag)^\mu
\Lambda^\nu]_{mn}~,\ee
where we used the identities like $\s_a e_\nu=i\eta_{\nu\lambda
a}e_\lambda$ and $\bar e_\mu e_\nu =
\delta_{\mu\nu}+i{\bar\eta}_{\mu\nu a} \s_a$. (Note that, in our
notation, $e_\mu \bar e_\nu = \delta_{\mu\nu}+i{\eta}_{\mu\nu
a}\s_a$). Then one finds that the matrix $\cal A$ should satisfy
the equation
\be\label{Aeq} -[\Omega^\mu,[\Omega^\mu,{\cal
A}]]-\{(\Lambda^\dag)^\mu \Lambda^\mu ,{\cal A}\}= -i\eta_{\mu\nu
a}\phi_0^a (\Lambda^\dag)^\mu \Lambda^\nu~,\ee
which is precisely what the authors of Refs.~\cite{dkm,kms,dhkm}
obtained as the condition for $\cal A$.

The 't Hooft $\kappa$ instanton solution \cite{thooft}
\be\label{thooft} A_\mu(x)= -{\bar\eta}_{\mu\nu
a}\frac{\s_a}{2i}\p_\nu \log \tilde\Pi(x)~, ~~~\tilde\Pi
(x)=1+\sum_{m=1}^\kappa \frac{\rho^2_m}{|x-z_m|^2} \ee
can be obtained from the simple canonical ADHM data: explicitly,
we here have ($z_m \equiv z^\mu_m e_\mu$)
\be B = \left(\begin{array}{ccc} \rho_1 & \cdots & \rho_\kappa \\
\hline z_1 & ~ & 0 \\ ~ & \ddots & ~ \\ 0 & ~ & z_\kappa
\end{array}\right)~,~~C=\left(\begin{array}{ccc} 0 & \cdots & 0 \\
\hline 1 & ~ & 0 \\ ~ & \ddots & ~ \\ 0 & ~ & 1
\end{array}\right)~,~~v(x)=\frac1{\sqrt{\tilde\Pi(x)}}\left(\begin{array}{c}
1 \\ \frac{\rho_1(x-z_1)}{|x-z_1|^2} \\ \vdots \\
 \frac{\rho_\kappa(x-z_\kappa)}{|x-z_\kappa|^2}\end{array}\right)~. \ee
In this case, a simple calculation shows that the inhomogeneous
term (\ref{inhom}) vanishes and hence ${\cal A}\equiv 0$. The
Higgs solution to the covariant Laplace equation is thus given by
\cite{seok}
\be\label{thphi} \phi(x)=\bar v_0 (x)\phi_0 v_0(x) =
\frac1{\tilde\Pi(x)}\phi_0~.\ee
But, with the JNR $\kappa$ instanton solution (which is described
more simply using noncanonical ADHM data), one finds ${\cal A}\neq
0$ if $\kappa >1$ and cannot expect this sort of simple solutions
any longer. For this JNR case, see Sec.~3 for detailed
discussions.

General $SU(2)$ dyonic instantons can be described by the ADHM
construction for $A_\mu(x)$ and the corresponding Higgs solution
(\ref{main2}). Using these solutions, we can evaluate the electric
charge $Q_e$ (given by (\ref{charge})) explicitly. Relegating the
details of calculation to Appendix B, we here give the final
result only:
\be\label{adhmchg} Q_e =-\frac{4\pi^2}{\rm v}{\rm tr}\{ \phi_0^2
V^\dag B(C^\dag C)^{-1}B^\dag V + 2\phi_0 V^\dag B{\cal A} B^\dag
V\} ~.\ee
(Here we invoked matrix notation in which (\ref{main2}) can be
written as $\phi=v^\dag V\phi_0 V^\dag v-2v^\dag C{\cal A}C^\dag
v$). If one takes the ADHM solution given in the canonical basis
and the corresponding Higgs solution (\ref{canon}) (the relevant
data are given in (\ref{canondata2})), (\ref{adhmchg}) reduces to
the simple form
\be\label{canchg} Q_e =2\pi^2 (\Lambda^\mu_i) \left[{\rm
v}\d_{\mu\nu}\d_{ij}-\frac{4\eta_{\mu\nu a}\phi_0^a}{\rm v}{\cal
A}_{ij}\right](\bar\Lambda^\nu_j) \ee
with the matrix $\cal A$ obtained by solving (\ref{Aeq}). An
equivalent expression to this result was found in
Ref.~\cite{tong}. For dyonic instantons given by the 't Hooft
configuration (\ref{thooft}) and the corresponding Higgs field
(\ref{thphi}), we find from (\ref{adhmchg}) or (\ref{canchg}) the
value $Q_e = 2\pi^2 {\rm v} (\rho_1^2 +\cdots + \rho_\kappa^2)$,
which is positive as ${\rm v}>0$. The above electric charge should
also be positive.

\section{Higgs Configurations with the Jackiw-Nohl-Rebbi
Instantons}

The Higgs solution (\ref{thphi}), obtained in the 't Hooft
instanton background, has $\kappa$ isolated zeroes at instanton
positions $x=z_1,\cdots,z_\kappa$; as argued in Ref.~\cite{seok},
these solutions describe collapsed supertubes connecting
D4-branes. To see any indication of supertubes with finite size
which connect two D4-branes, it was suggested \cite{seok} to study
the Higgs configuration in the JNR background \cite{jnr}. The
latter is obtained by modifying (\ref{thooft}) into the form
\be\label{jnr} A_\mu(x)= -{\bar\eta}_{\mu\nu
a}\frac{\s_a}{2i}\p_\nu \log \Pi(x)~, ~~~\Pi (x)=\sum_{l=0}^\kappa
\frac{\rho^2_l}{|x-z_l|^2}~. \ee
This $\kappa$ instanton configuration, with four more physically
relevant parameters than the 't Hooft instanton configuration
(\ref{thooft})\footnote{When the $\kappa +1$ points
$z_0,z_1,\cdots,z_\kappa$ lie on a circle or a line, there is a
residual gauge degree of freedom to reduce the number of
parameters to $5\kappa +3$, three more than the 't Hooft case
\cite{jnr}.}, can be obtained from the noncanonical ADHM data
\cite{cori,seok}
\be\label{jnrdata} B = \left(\begin{array}{cccc}
-\frac{\rho_1}{\rho_0}z_0 & -\frac{\rho_2}{\rho_0}z_0 & \cdots &
-\frac{\rho_\kappa}{\rho_0} z_0 \\
\hline z_1 & ~ & ~ & 0 \\ ~ & z_2 & ~ & ~ \\ ~ & ~ & \ddots & ~
\\ 0 & ~ & ~ & z_\kappa
\end{array}\right)~,~~C=\left(\begin{array}{cccc}
-\frac{\rho_1}{\rho_0} &  -\frac{\rho_2}{\rho_0}
 & \cdots & -\frac{\rho_\kappa}{\rho_0} \\
\hline 1 & ~ & ~ & 0 \\ ~ & 1 & ~ & ~ \\ ~ & ~ & \ddots & ~ \\ 0 &
~ & ~ & 1
\end{array}\right)~. \ee
With these data the quaternionic column vector $v(x)$ is readily
found to be
\be\label{jnrv} v(x)=\frac1{\sqrt{\Pi(x)}} \begin{pmatrix}
\frac{\rho_0(x-z_0)}{|x-z_0|^2} \\ \frac{\rho_1(x-z_1)}{|x-z_1|^2}
\\ \vdots \\ \frac{\rho_\kappa(x-z_\kappa)}{|x-z_\kappa |^2}
\end{pmatrix}~,\ee
and hence, for the quantity $h_0(\hat y)=Vg(\hat y)$ (see
(\ref{asym1})), we have
\be\label{jnrasym} g(\hat y) = \hat y~,~~~V=\frac1{\sqrt
S}\begin{pmatrix} \rho_0
\\ \rho_1 \\ \vdots \\ \rho_\kappa \end{pmatrix}~,~~~\left(S\equiv
\sum_{l=0}^\kappa \rho_l^2\right)~.\ee

The Higgs solution in the JNR instanton background can be
calculated by using the above data with our formula (\ref{main2}).
The first contribution in (\ref{main2}), denoted $\phi_{\rm
I}(x)$, is simple:
\ba \phi_{\rm I}(x) &\equiv& v^\dag(x) V\phi_0 V^\dag v(x)
\nn\\&=& \frac1{S\:\Pi(x)}\bar{\cal F}(x)\phi_0 {\cal
F}(x)~,~~~\left({\cal F}(x)\equiv \sum_{l=0}^\kappa
\frac{\rho_l^2(x-z_l)}{|x-z_l|^2}\right)\ea
where $S$ is a constant factor defined in (\ref{jnrasym}). More
explicitly, we can write this as
\be\label{phi1} \phi_{\rm I}(x) =
-\frac1{S\:\Pi(x)}\frac{\s_a}{2i}\left[ \sum_{ll'=0}^\kappa
\rho_l^2 \rho_{l'}^2\: X^\mu_l X^\nu_{l'}\: \bar\eta_{\mu\lambda
a} \eta_{\nu\lambda b}\right]\phi_0^b  ~,\ee
where $X^\mu_l(x)\equiv (x-z_l)^\mu/|x-z_l|^2$
($l=0,1,\cdots,\kappa$). In the present case the second term in
(\ref{main2}) also contributes to the Higgs configuration if
$\kappa >1$, since we now find, for the inhomogeneous term in
(\ref{A}),
\be\label{jnrA} {\rm tr}(B^\dag V\phi_0 V^\dag B)_{mn} =-\frac1{S}
\rho_m \rho_n \eta_{\lambda\d b} \phi_0^b (z_m -z_0)^\lambda
(z_n-z_0)^\d \ee
and so some ${\cal A}_{ij}$ should not be zero. By substituting
the ADHM data (\ref{jnrdata}) in (\ref{L}), we here obtain
\ba L_{mn,rs} &=& \left[ -(z_m-z_n)^2 \d_{mr}\d_{ns}
-\frac{\rho_n\rho_s}{\rho_0^2}(z_m-z_0)^2 \d_{mr}
-\frac{\rho_m\rho_r}{\rho_0^2}(z_n-z_0)^2
\d_{ns}\right]-[m\leftrightarrow n]\nn\\ &=&
-\rho_m\rho_n\rho_r\rho_s\left[
\frac{(z_m-z_n)^2}{\rho_m^2\rho_n^2}
(\d_{mr}\d_{ns}-\d_{ms}\d_{nr}) +\frac{(z_m-z_0)^2
}{\rho_0^2\rho_m^2}(\d_{mr}-\d_{ms})\right.\nn\\&~&\label{jnrL}\left.
+\frac{(z_n-z_0)^2}{\rho_0^2\rho_n^2} (\d_{ns}-\d_{nr})\right]~.
\ea
Now, based on the expressions (\ref{jnrA}) and (\ref{jnrL}), our
linear simultaneous equations (\ref{A}) for the present case can
be recast into the form
\be\label{39} [\tilde
z_{mn}^2(\d_{mr}\d_{ns}-\d_{ms}\d_{nr})+\tilde
z_{m0}^2(\d_{mr}-\d_{ms})+\tilde z_{n0}^2 (\d_{ns} -
\d_{nr})]{\tilde{\cal A}}_{rs} = \frac1{S} \eta_{\lambda\d b}
\phi_0^b (z_0^\lambda z_m^\d + z_m^\lambda z_n^\d +z_n^\lambda
z_0^\d)~,\ee
where we defined
\be {\tilde{\cal A}}_{rs} \equiv \rho_r \rho_s {\cal
A}_{rs}~,~~~\tilde z_{kl}^2 \equiv
\frac{(z_k-z_l)^2}{\rho_k^2\rho_l^2}~,~~~(k,l=0,1,\cdots,\kappa)~.\ee
{}From the very structure of the linear equations (\ref{39}), we
will also define the corresponding inverse matrix $\tilde
f_{rs,mn}$ so that we may write
\be {\tilde{\cal A}}_{rs} = -\frac1{2S}\eta_{\lambda\d b}\phi_0^b
\sum_{mn=1}^\kappa \tilde f_{rs,mn} (z_0^\lambda z_m^\d +
z_m^\lambda z_n^\d +z_n^\lambda z_0^\d)~.\ee
At the same time, since we find from the expressions in
(\ref{jnrdata}) and (\ref{jnrv})
\be  (v^\dag(x)C)_i (C^\dag v(x))_j - (v^\dag(x)C)_j (C^\dag
v(x))_i = \frac{\rho_i\rho_j}{\Pi(x)}
(X^\mu_i-X^\mu_0)(X^\nu_j-X^\nu_0)\bar e_\mu e_\nu
-(i\leftrightarrow j)~,\ee
(the $X$'s here denote the same variables introduced in
(\ref{phi1})), the second term of (\ref{main2}) can be represented
by the form
\ba\phi_{\rm II}(x) &\equiv& -\sum_{ij=1}^\kappa [(v^\dag(x)C)_i
(C^\dag v(x))_j - (v^\dag(x)C)_j (C^\dag v(x))_i]{\cal
A}_{ij}\nn\\&=&-\frac2{S\:\Pi(x)}\frac{\s_a}{2i}\left[\sum_{ijmn=1}^\kappa
(X_0^\mu X_i^\nu +X_i^\mu X_j^\nu +X_j^\mu X_0^\nu
)\bar\eta_{\mu\nu a}\eta_{\lambda\d b} \tilde f_{ij,mn}
(z_0^\lambda z_m^\d + z_m^\lambda z_n^\d +z_n^\lambda
z_0^\d)\right]\phi_0^b~.\nn\\ \label{phi2} ~ \ea
Based on (\ref{phi1}) and (\ref{phi2}), the full Higgs
configuration in the JNR instanton background is
\ba\phi(x) &=&
-\frac1{S\:\Pi(x)}\frac{\s_a}{2i}\Bigg[\sum_{ll'=0}^\kappa
\rho_l^2\rho_{l'}^2 X_l^\mu X_{l'}^\nu \bar\eta_{\mu\lambda
a}\eta_{\nu\lambda b}\nn\\&~& +2\sum_{ijmn=1}^\kappa (X_0^\mu
X_i^\nu +X_i^\mu X_j^\nu +X_j^\mu X_0^\nu )\bar\eta_{\mu\nu
a}\eta_{\lambda\d b} \tilde f_{ij,mn} (z_0^\lambda z_m^\d +
z_m^\lambda z_n^\d +z_n^\lambda z_0^\d)\Bigg]\phi_0^b\nn\\
\label{full} ~ \ea
with the constants $\tilde f_{ij,mn}$ ($=-\tilde f_{ji,mn}=-\tilde
f_{ij,nm}$) determined  by inverting the linear inhomogeneous
equations (\ref{39}).

Based on (\ref{full}), we will now produce explicit Higgs
configurations appropriate to dyonic instantons with some small
instanton number $\kappa$. In the $\kappa=1$ JNR instanton
background the corresponding Higgs configuration is very simple,
being given by
\be \phi(x) = -\frac{1
}{S\:\Pi(x)}\frac{\s_a}{2i}\left[\bar\eta_{\mu\lambda
a}\eta_{\nu\lambda b}(\rho_0^2 X_0+\rho_1^2 X_1)^\mu (\rho_0^2
X_0+\rho_1^2 X_1)^\nu \right]\phi_0^b\ee
together with obvious identifications for $S$ and $\Pi(x)$, i.e.,
$S=\rho_0^2 +\rho_1^2$ and $\Pi(x)=\frac{\rho_0^2}{|x-z_0|^2}
+\frac{\rho_1^2}{|x-z_1|^2}$. It should be noted, however, that
all $\kappa=1$ dyonic instanton solutions are gauge-equivalent to
corresponding 't Hooft-type dyonic instanton solutions. With
$\kappa=2$, nonzero elements of the matrix $\tilde f^{-1}$
--- the matrix in terms of which the left hand side of (\ref{39})
can be written $-(\tilde f^{-1})_{mn,rs} {\tilde{\cal A}}_{rs}$
(with $(\tilde f^{-1})_{mn,rs}\tilde
f_{rs,ij}=\d_{mi}\d_{nj}-\d_{mj}\d_{ni}$ )
--- are restricted to
\be\label{finv} (\tilde f^{-1})_{12,12}=-(\tilde
f^{-1})_{21,12}=-(\tilde f^{-1})_{12,21}=(\tilde
f^{-1})_{21,21}=-(\tilde z_{01}^2 +\tilde z_{12}^2 + \tilde
z_{20}^2)~,\ee
and therefore we find, for nonzero elements of the matrix $\tilde
f$,
\be\label{f1212} \tilde f_{12,12}=-\tilde f_{21,12}=-\tilde
f_{12,21}=\tilde f_{21,21}=-\frac1{2(\tilde z_{01}^2 +\tilde
z_{12}^2 + \tilde z_{20}^2)}~.\ee
Using this result in (\ref{full}), we obtain the appropriate Higgs
configuration
\ba \phi(x)&=&
-\frac1{S\:\Pi(x)}\frac{\s_a}{2i}\Bigg[\bar\eta_{\mu\lambda
a}\eta_{\nu\lambda b}(\rho_0^2 X_0 +\rho_1^2 X_1 +\rho_2^2
X_2)^\mu (\rho_0^2 X_0 +\rho_1^2 X_1 +\rho_2^2 X_2)^\nu
\nn\\\label{2Higgs}&~& -4\:\bar\eta_{\mu\nu a}\eta_{\lambda\d
b}(X^\mu_0 X^\nu_1 +X^\mu_1 X^\nu_2 +X^\mu_2 X^\nu_0)
\frac{(z_0^\lambda z_1^\d +z_1^\lambda z_2^\d +z_2^\lambda
z_0^\d)}{\tilde z_{01}^2 +\tilde z_{12}^2 + \tilde
z_{20}^2}\Bigg]\phi_0^b~. \ea
The result equivalent to this expression was obtained earlier in
Ref.~\cite{seok}. Notice that this form exhibits a full symmetry
under the interchange of three sets of JNR parameters,
$(\rho_0,z^\mu_0)$, $(\rho_1,z^\mu_1)$ and $(\rho_2,z^\mu_2)$.

For $\kappa=3$ it is convenient to define ${\tilde{\cal A}}_i$
($i=1,2,3$) by ${\tilde{\cal A}}_{rs}=\epsilon_{rsi}{\tilde{\cal
A}}_i$, so that (\ref{39}) may be recast as
\be\label{319} \begin{pmatrix} \tilde z_{23}^2+\tilde
z_{20}^2+\tilde z_{30}^2 & -\tilde z_{30}^2 & -\tilde z_{20}^2 \\
-\tilde z_{30}^2 & \tilde z_{31}^2+\tilde z_{30}^2+\tilde z_{10}^2
& -\tilde z_{10}^2 \\ -\tilde z_{20}^2 & -\tilde z_{10}^2 & \tilde
z_{12}^2+\tilde z_{10}^2+\tilde z_{20}^2
\end{pmatrix} \begin{pmatrix} {\tilde{\cal A}}_1 \\ {\tilde{\cal A}}_2
\\ {\tilde{\cal A}}_3 \end{pmatrix}=\frac1{S}\eta_{\lambda\d b} \phi_0^b
\begin{pmatrix}z_0^\lambda z_2^\d +z_2^\lambda z_3^\d +z_3^\lambda
z_0^\d \\ z_0^\lambda z_3^\d +z_3^\lambda z_1^\d +z_1^\lambda
z_0^\d \\ z_0^\lambda z_1^\d +z_1^\lambda z_2^\d +z_2^\lambda
z_0^\d \end{pmatrix}~. \ee
After somewhat tedious algebra, one can determine $\tilde{\cal
A}_i$ and hence the expression for $\phi_{\rm II}(x)$ (see
(\ref{phi2})) also. With some rearrangements one can then write
the explicit Higgs configuration that go with $\kappa=3$ JNR
instanton in the form
\ba \phi(x)&=&
-\frac1{S\:\Pi(x)}\frac{\s_a}{2i}\Bigg[\bar\eta_{\mu\lambda
a}\eta_{\nu\lambda b}\Big(\sum_{l=0}^3 \rho_l^2 X^\mu_l
\Big)\Big(\sum_{l'=0}^3 \rho_{l'}^2 X^\nu_{l'}\Big)\nn
\\  &~& -\frac2{D^{(3)}}\bar\eta_{\mu\nu a} \eta_{\lambda\d
b}\Big\{ \sum_{ll'=0}^3 X_{l}^\mu
X_{l'}^\nu\big[P^{(3)}_{ll'}z^\lambda_l
z^\d_{l'}+2\sum_{k\neq(l,l')}^3 P^{(3)}_{ll'k}z^\lambda_{l'}z^\d_k
+\sum_{kk'\neq(l,l')}^3 P^{(3)}_{ll',kk'}z^\lambda_k z^\d_{k'}
\big]
 \Big\}\Bigg]\phi_0^b ~,\nn\\ \label{3Higgs} ~\ea
where $D^{(3)}$, the determinant of the $3\times 3$ matrix
appearing in the left hand side of (\ref{319}), is given by
\ba D^{(3)} &=& \tilde z^2_{12} \tilde z^2_{20} \tilde z^2_{10}
+\tilde z^2_{31} \tilde z^2_{30} \tilde z^2_{10} +\tilde z^2_{23}
\tilde z^2_{30} \tilde z^2_{20} +\tilde z^2_{12} \tilde z^2_{23}
\tilde z^2_{31}\nn\\\label{D}&~&~~ +\tilde z^2_{31} \tilde
z^2_{23} \tilde z^2_{20} +\tilde z^2_{12} \tilde z^2_{23} \tilde
z^2_{30} +\tilde z^2_{23} \tilde z^2_{31} \tilde z^2_{10} +\tilde
z^2_{12} \tilde z^2_{31} \tilde z^2_{30}\\&~&~~ +\tilde z^2_{23}
\tilde z^2_{30} \tilde z^2_{10} +\tilde z^2_{31} \tilde z^2_{30}
\tilde z^2_{20} +\tilde z^2_{10} \tilde z^2_{12} \tilde
z^2_{23} +\tilde z^2_{31} \tilde z^2_{12} \tilde z^2_{20}\nn\\
&~&~~ +\tilde z^2_{10} \tilde z^2_{20} \tilde z^2_{23} +\tilde
z^2_{12} \tilde z^2_{20} \tilde z^2_{30} +\tilde z^2_{31} \tilde
z^2_{10} \tilde z^2_{20} +\tilde z^2_{12} \tilde z^2_{10} \tilde
z^2_{30}~,\nn \ea
and $P^{(3)}_{ll'}$, $P^{(3)}_{ll'k}$, $P^{(3)}_{ll',kk'}$ (with
any of the indices $l,l',k$ and $k'$ taking values among 0, 1, 2,
3) are some quadratic polynomials of the $\tilde z^2$'s precise
form of which we will specify below.

Observe that the determinant $D^{(3)}$ is a cubic polynomial of
the six elements $\tilde z^2_{01}$, $\tilde z^2_{02}$, $\tilde
z^2_{03}$, $\tilde z^2_{12}$, $\tilde z^2_{13}$ and $\tilde
z^2_{23}$, satisfying the conditions that (i) no element appears
more than once in each monomial, (ii) no given index
$l$($=0,1,2,3$) appears more than twice in each monomial, and
(iii) the full sum exhibits symmetry under the interchange of four
JNR parameters $(\rho_0,z_0^\mu)$, $(\rho_1,z_1^\mu)$,
$(\rho_2,z_2^\mu)$ and $(\rho_3,z_3^\mu)$. Then the polynomials
$P^{(3)}_{ll'}$, each given by the sum of eight quadratic
monomials of the $\tilde z^2$'s, are related to $D^{(3)}$ by
\be P^{(3)}_{ll'}=\frac{\p D^{(3)}}{\p \tilde
z^2_{ll'}}~~(=P^{(3)}_{l'l})~.\ee
Now let $P^{(3)}_{ll'}\cap P^{(3)}_{kk'}$ denote the sum of all
monomials which make simultaneous appearance in both polynomials
$P^{(3)}_{ll'}$ and $P^{(3)}_{kk'}$. We then find that the
polynomials $P^{(3)}_{ll'k}$ above, each corresponding to the sum
of four quadratic monomials of the $\tilde z^2$'s, can be
identified with
\be P^{(3)}_{ll'k} = P^{(3)}_{ll'}\cap P^{(3)}_{l'k}~,\ee
while the polynomials $P^{(3)}_{ll',kk'}$, in the last piece of
(\ref{3Higgs}), equal
\be P^{(3)}_{ll',kk'}= \tilde z^2_{lk} \tilde z^2_{l'k'}
\frac{\p}{\p \tilde z^2_{lk}}\frac{\p}{\p \tilde z^2_{l'k'}}
(P^{(3)}_{ll'}\cap P^{(3)}_{kk'})~.\ee
Explicitly, we have
\ba P^{(3)}_{01} &=&  {{\tilde z_{03}}}^2\,{{\tilde z_{12}}}^2 +
      {{\tilde z_{12}}}^2\,{{\tilde z_{23}}}^2 +
       {{\tilde z_{32}}}^2\,{{\tilde z_{{20}}}}^2 + {{\tilde z_{02}}}^2\,
       {{\tilde z_{{21}}}}^2 +
       {{\tilde z_{02}}}^2\,{{\tilde z_{{13}}}}^2 + {{\tilde z_{23}}}^2\,
       {{\tilde z_{{30}}}}^2 +
       {{\tilde z_{03}}}^2\,{{\tilde z_{{31}}}}^2 + {{\tilde z_{13}}}^2\,
       {{\tilde z_{{32}}}}^2~,   \nn\\
P^{(3)}_{12} &=& {{\tilde z_{10}}}^2\,{{\tilde z_{23}}}^2 +
      {{\tilde z_{13}}}^2\,{{\tilde z_{32}}}^2 +
       {{\tilde z_{23}}}^2\,{{\tilde z_{{30}}}}^2 + {{\tilde z_{03}}}^2\,
       {{\tilde z_{{31}}}}^2+
       {{\tilde z_{13}}}^2\,{{\tilde z_{{20}}}}^2 + {{\tilde z_{10}}}^2\,
       {{\tilde z_{{03}}}}^2 +
       {{\tilde z_{30}}}^2\,{{\tilde z_{{02}}}}^2 + {{\tilde z_{20}}}^2\,
       {{\tilde z_{{01}}}}^2~,  \nn\\
P^{(3)}_{012} &=& P^{(3)}_{01}\cap P^{(3)}_{12} ~=~  {{\tilde
z_{13}}}^2\,{{\tilde z_{20}}}^2 +
       {{\tilde z_{23}}}^2\,{{\tilde z_{{30}}}}^2 +
       {{\tilde z_{03}}}^2\,{{\tilde z_{{31}}}}^2 + {{\tilde z_{13}}}^2\,
       {{\tilde z_{{32}}}}^2~, \\
P^{(3)}_{0123} &=& \tilde z^2_{02}\tilde z^2_{13}~,\nn\ea
etc. We remark that the resulting Higgs configuration has full
symmetry under the exchange of four sets of JNR parameters.

Beyond $\kappa=3$ the algebra involved in solving (\ref{39})
becomes very complicated. But, inferring from the detailed
analysis we performed for the case of $\kappa=4$, the Higgs
configuration for $\kappa\geq 4$ appears to be described by a
direct extension of our $\kappa=3$ formula (\ref{3Higgs}), i.e.,
by using now the related determinant $D^{(\kappa)}$ (associated
with the $\frac{\kappa(\kappa-1)}2\times\frac{\kappa(\kappa-1)}2$
matrix, formed from the coefficients multiplying the $\tilde{\cal
A}$'s in (\ref{39})) and the $\tilde z^2$-dependent quantities
$P^{(\kappa)}_{ll'}$, $P^{(\kappa)}_{ll'k}$ and
$P^{(\kappa)}_{ll',kk'}$ given by
\ba P^{(\kappa)}_{ll'}=\frac{\p D^{(\kappa)}}{\p \tilde z^2_{ll'}}~,~~
P^{(\kappa)}_{ll'k} = P^{(\kappa)}_{ll'}\cap P^{(\kappa)}_{l'k}~, \nn\\
P^{(\kappa)}_{ll',kk'}= \tilde z^2_{lk} \tilde z^2_{l'k'}
\frac{\p}{\p \tilde z^2_{lk}}\frac{\p}{\p \tilde z^2_{l'k'}}
(P^{(\kappa)}_{ll'}\cap P^{(\kappa)}_{kk'})~. \ea
Actually, even for $\kappa=2$, the corresponding form with
$D^{(2)}=\tilde z^2_{01}+\tilde z^2_{12}+\tilde z^2_{20}$,
$P^{(2)}_{ll'}=1$ for $l\neq l'$, $P^{(2)}_{ll'k}=1$ for
$k\neq(l,l')$, and $P^{(2)}_{ll',kk'}\equiv 0$ reproduces the
expression (\ref{2Higgs}) exactly. For $\kappa=4$, $D^{(4)}$ --- a
sixth-order polynomial satisfying the restrictions that (i) no
element appears more than once in each monomial, (ii) no given
index $l$($=0,1,2,3,4$) appears more than three times in each
monomial, (iii) no monomial of the type $\tilde z^2_{02}\tilde
z^2_{03}\tilde z^2_{04}\tilde z^2_{12}\tilde z^2_{13}\tilde
z^2_{14}$ is kept, and (iv) the full sum exhibits symmetry under
the interchange of five sets of JNR parameters. In this case we
verified that the full Higgs configuration is represented by a
direct generalization of (\ref{3Higgs}); no additional term is
needed, whatsoever. We conjecture that this be the case for
$\kappa\geq 5$ also. Accepting this, the full Higgs configurations
appropriate to JNR dyonic instantons follow only if the suitable
expression for the determinant associated with our linear
equations (\ref{39}) has been evaluated. The symmetry of the
configuration under the exchange of JNR parameters will be
automatic.

The electric charge for JNR dyonic instantons can be computed
using our formula (\ref{adhmchg}). Especially, for the
$\kappa=1,2$ and 3 cases, one obtains following values:
\ba\label{1charge} \kappa=1 &:& Q_e = 2\pi^2 {\rm v}
\frac{\rho_0^2\rho_1^2}{(\rho_0^2+\rho_1^2)^2}(z_1-z_0)^2~,\\\label{2charge}
\kappa=2 &:& Q_e =\frac{2\pi^2}{{\rm
v}(\rho_0^2+\rho_1^2+\rho_2^2)^2} \Bigg\{ {\rm v}^2 [\rho_0^2
\rho_1^2 (z_0-z_1)^2 + \rho_1^2 \rho_2^2 (z_1-z_2)^2 + \rho_2^2
\rho_0^2 (z_2-z_0)^2]\nn\\&~&~~~~~~~~ -\frac{4[\eta_{\mu\nu a}
\phi_0^a (z_0^\mu z_1^\nu +z_1^\mu z_2^\nu +z_2^\mu
z_0^\nu)]^2}{\tilde z_{01}^2 +\tilde z_{12}^2
+\tilde z_{20}^2}\Bigg\}~,\\
\label{3charge} \kappa=3 &:& Q_e =\frac{2\pi^2}{{\rm
v}\left(\sum_{l=0}^3\rho_l^2\right)^2}\Bigg\{{\rm v}^2[\rho_0^2
\rho_1^2 (z_0-z_1)^2 + \rho_0^2 \rho_2^2 (z_0-z_2)^2 + \rho_0^2
\rho_3^2 (z_0-z_3)^2 \nn\\ &~&~~~ +\rho_1^2 \rho_2^2 (z_1-z_2)^2 +
\rho_1^2 \rho_3^2 (z_1-z_3)^2 + \rho_2^2 \rho_3^2
(z_2-z_3)^2]\nn\\&~&~~~
-\frac{4P^{(3)}_{12}}{D^{(3)}}[\eta_{\mu\nu a} \phi_0^a (z_0^\mu
z_1^\nu +z_1^\mu z_2^\nu +z_2^\mu z_0^\nu)]^2
-\frac{4P^{(3)}_{23}}{D^{(3)}}[\eta_{\mu\nu a} \phi_0^a (z_0^\mu
z_2^\nu +z_2^\mu z_3^\nu +z_3^\mu
z_0^\nu)]^2\nn\\&~&~~~-\frac{4P^{(3)}_{31}}{D^{(3)}}[\eta_{\mu\nu
a} \phi_0^a (z_0^\mu z_3^\nu +z_3^\mu z_1^\nu +z_1^\mu
z_0^\nu)]^2\nn\\&~&~~~-\frac{8P^{(3)}_{123}}{D^{(3)}}[\eta_{\mu\nu
a} \phi_0^a (z_0^\mu z_1^\nu +z_1^\mu z_2^\nu +z_2^\mu
z_0^\nu)][\eta_{\lambda\d b} \phi_0^b (z_0^\lambda z_2^\d
+z_2^\lambda z_3^\d +z_3^\lambda z_0^\d)]\nn
\\&~&~~~-\frac{8P^{(3)}_{213}}{D^{(3)}}[\eta_{\mu\nu a} \phi_0^a
(z_0^\mu z_1^\nu +z_1^\mu z_2^\nu +z_2^\mu
z_0^\nu)][\eta_{\lambda\d b} \phi_0^b (z_0^\lambda z_3^\d
+z_3^\lambda z_1^\d +z_1^\lambda z_0^\d)]\nn
\\&~&~~~-\frac{8P^{(3)}_{231}}{D^{(3)}}[\eta_{\mu\nu a} \phi_0^a
(z_0^\mu z_2^\nu +z_2^\mu z_3^\nu +z_3^\mu
z_0^\nu)][\eta_{\lambda\d b} \phi_0^b (z_0^\lambda z_3^\d
+z_3^\lambda z_1^\d +z_1^\lambda z_0^\d)]\Bigg\} ~.\ea
The result (\ref{1charge}) for $\kappa=1$ JNR dyonic instanton is
easily understood, based on the facts that (i) the configuration
with JNR parameters $(\rho_0,z_0^\mu)$ and $(\rho_1,z_1^\mu)$ is
gauge-equivalent to the 't Hooft-type dyonic instanton with size
$\rho = \frac{\rho_0\rho_1}{\rho_0^2+\rho_1^2}|z_1-z_0|$ and
position $z^\mu=\frac{\rho_1^2 z_0^\mu +\rho_0^2 z_1^\mu}{\rho_0^2
+\rho_1^2}$, and (ii) for the latter we obtained the result
$Q_e=2\pi^2 {\rm v} \:\rho^2$ already. Our expression
(\ref{2charge}) coincides with the result obtained earlier
\cite{seok}. As regars our result (\ref{3charge}) giving the value
for $\kappa=3$, we have checked explicitly that the given
expression exhibits the full symmetry under the interchange of
related four JNR parameters, despite its partly asymmetric
appearance.
 We also remark
that both the results (\ref{2charge}) and (\ref{3charge}) reduce to
the corresponding electric charge values of 't Hooft-type dyonic
instantons if we take the limit $\rho_0^2\rightarrow\infty$,
$z_0^2\rightarrow\infty$ with the ratio $\rho_0^2/z_0^2$ held to 1.

\section{Higgs Zero Locus and Connection to Supertubes}

In the previous section, we obtained a very explicit form of Higgs
configuration (\ref{3Higgs}) for JNR-type dyonic instanton with
$\kappa =3$, and reported an observation that the same form for
the Higgs solutions also goes for the $\kappa =4$ case. Based on
this observation, we propose the following form of Higgs
configuration for an arbitrary topological charge $\kappa$:
\ba \phi(x)&=&
-\frac1{S\:\Pi(x)}\frac{\s_a}{2i}\Bigg[\bar\eta_{\mu\lambda
a}\eta_{\nu\lambda b}\Big(\sum_{l=0}^\kappa \rho_l^2 X^\mu_l
\Big)\Big(\sum_{l'=0}^\kappa \rho_{l'}^2 X^\nu_{l'}\Big)\nn
\\  &~& -\frac2{D^{(\kappa)}}\bar\eta_{\mu\nu a} \eta_{\lambda\d
b}\Bigg\{ \sum_{ll'=0}^\kappa X_{l}^\mu
X_{l'}^\nu\bigg[P^{(\kappa)}_{ll'}z^\lambda_l
z^\d_{l'}+2\sum_{k\neq(l,l')}^\kappa
P^{(\kappa)}_{ll'k}z^\lambda_{l'}z^\d_k
+\sum_{kk'\neq(l,l')}^\kappa P^{(\kappa)}_{ll',kk'}z^\lambda_k
z^\d_{k'} \bigg]
 \Bigg\}\Bigg]\phi_0^b ~.\nn\\ \label{kHiggs} ~\ea
With this expression, we here present some analysis on the zeroes of
the above Higgs field $\phi(x)$ . For this purpose it is convenient
to rewrite (\ref{kHiggs}) as follows:
\begin{eqnarray}
  \phi(x)&\equiv& \frac{1}{S\:\Pi(x)} \frac{\sigma_a}{2i}  M^{ab}
\phi_0^b\nn\\\nn &=& \frac{1}{S\:\Pi(x)} \frac{\sigma_a}{2i}
\left[ 2 {\cal F}^a {\cal F}^b -({\cal F}^c {\cal F}^c  - {\cal
F}^0{\cal F}^0 )\delta ^{ab} +2 \epsilon^{abc}{\cal F}^c {\cal
F}^0  +
\sum_{ll'=0}^{\kappa} (V_{ll'})_a (W_{ll'})_b\right]\phi_0^b~,\\
\label{new form higgs}~
\end{eqnarray}
where we have defined
\begin{eqnarray}\label{forces}
&& {\cal F}^\mu = \sum_{l=0}^{\kappa} {\cal F}_l^\mu~~,~~{\cal
F}_l^\mu = \rho_l^2 X_l^\mu = \frac{\rho_l^2
(x-z_l)^\mu}{|x-z_l|^2}~,
\\
&&(V_{ll'})_a = 2 \bar\eta_{\mu\nu a} X^\mu_l X^\nu_{l'}~,
 \\
&&(W_{ll'})_b = \frac{\eta_{\lambda\d
b}}{D^{(\kappa)}}\left[P^{(\kappa)}_{ll'}z^\lambda_l
z^\d_{l'}+2\sum_{k\neq(l,l')}^\kappa
P^{(\kappa)}_{ll'k}z^\lambda_{l'}z^\d_k +\sum_{kk'\neq(l,l')}^\kappa
P^{(\kappa)}_{ll',kk'}z^\lambda_k z^\d_{k'} \right]~.
\end{eqnarray}
Note that the Higgs field is linearly proportional to its
asymptotic value $\phi^a_0$. Thus the zero locus will be
independent of the magnitude of the asymptotic value. As we
increase its value, the corresponding electric charge also
increases linearly, counterbalancing the force which is trying to
shrink instantons. However, the zero locus will depend crucially
on the orientation of the asymptotic value $\phi^a_0$. Usually we
fix this quantity to be diagonal and change the orientation of the
instanton configuration, but here we may fix the instanton field
configuration and consider changing the orientation of the
asymptotic value for convenience. For Higgs fields to vanish, the
matrix $M=M^{ab}$ (defined in (\ref{new form higgs})) should have
an eigenvector with zero eigenvalue or $\det M=0$. As it is a
single equation on four coordinates $x^\mu$, the equation $\det
M=0$ defines a 3-dimensional hypersurface. This hypersurface
represents the collection of all zeroes of the Higgs field as we
change the orientation of its asymptotic value. There are two
parameters in choosing the orientation of the asymptotic Higgs
field, implying that there is 1-dimensional zero locus for a given
orientation as expected.

\begin{figure}[t]
\centering\psfrag{A}{A}\psfrag{B}{B}\psfrag{C}{C}
\psfrag{x1}{$x^1$}\psfrag{x2}{$x^2$}\psfrag{x3}{$x^3$}
\psfrag{x2=0}{$x^2=0$}\psfrag{x3=0}{$x^3=0$}
\includegraphics{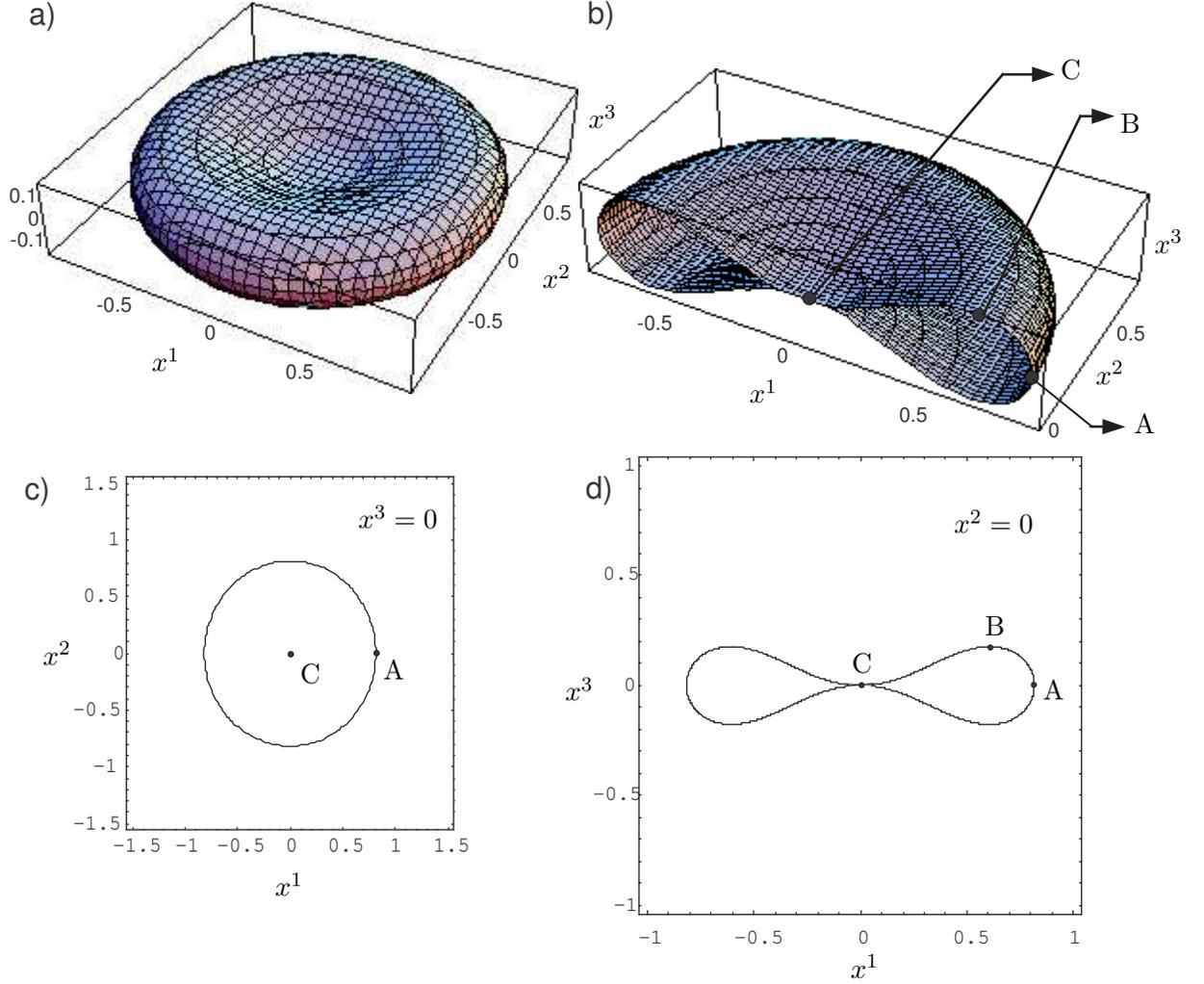}\caption{a) The
surface of vanishing determinant of $M$ at $x^0 =0$, with identical
size parameters $\rho_0 = \rho_1 = \rho_2 = 1$ and symmetric
position parameters $z_l$'s chosen at $z_0 = (-1,0)$, $z_1 =
(\frac{1}{2},\frac{\sqrt{3}}{2})$ and
$z_2=(\frac{1}{2},-\frac{\sqrt{3}}{2})$. b) The half of the surface
which shows the section $x^2=0$ of it. c) The section $x^3=0$ of the
surface. d) The section $x^2=0$ of the surface.}\label{gp1}
\end{figure}

\begin{figure}[t]
\centering\psfrag{x1}{$x^1$}\psfrag{x2}{$x^2$}\psfrag{x3}{$x^3$}\psfrag{x3=0}{$x^3=0$}
\includegraphics{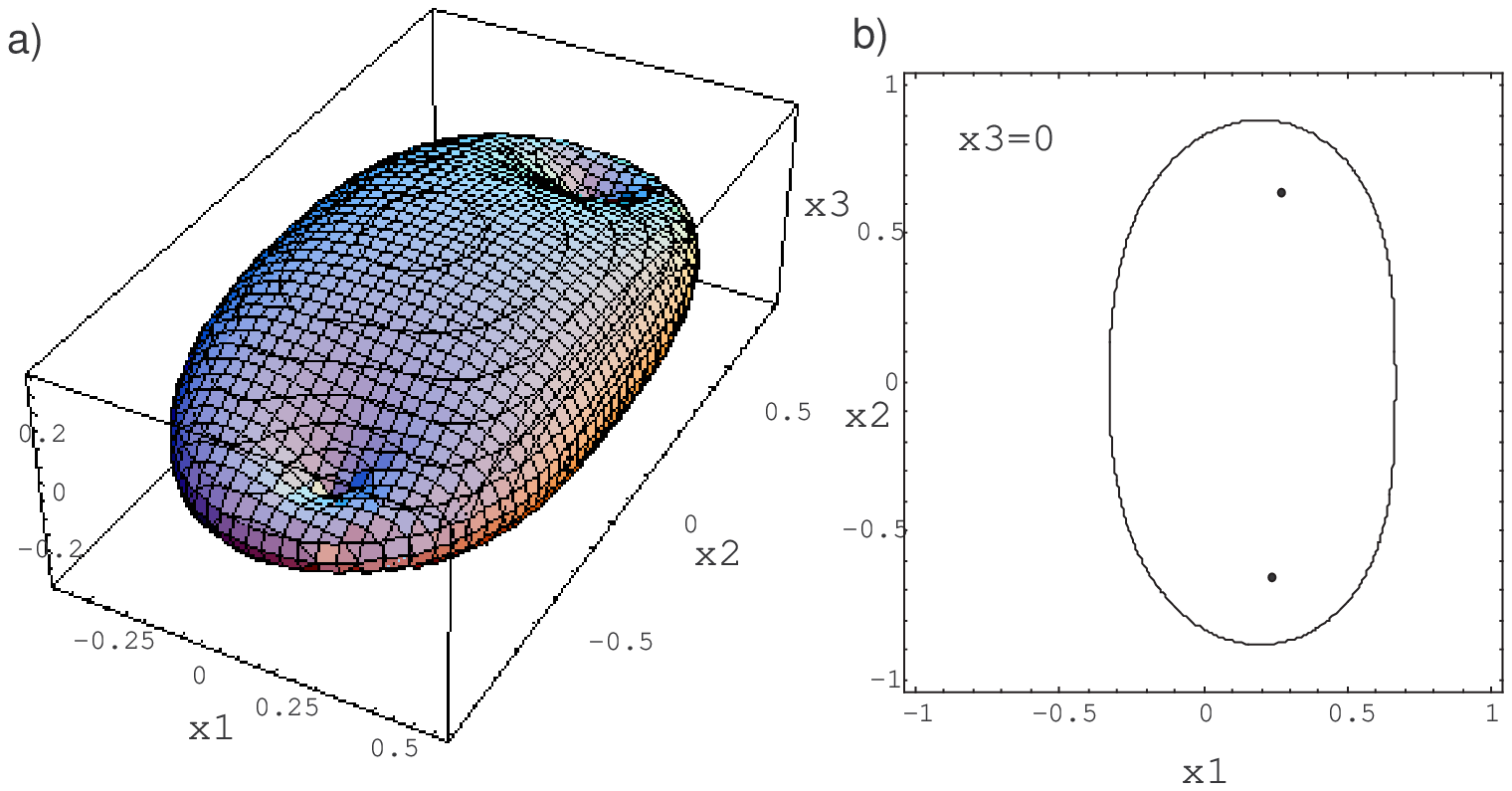}\caption{a) The surface $\det M=0$ at $x^0 =0$
with JNR parameters $\rho_0 =2, \rho_1 =\rho_2 =1, z_0 = (-1, 0),
z_1 = (\frac{1}{2}, \frac{\sqrt {3}}{2})$ and $z_2= (\frac{1}{2}
-\frac{\sqrt{3}}{2})$. b) The section $x^3=0$ of the
surface.}\label{gp2}
\end{figure}

We shall begin our analysis of the Higgs zeroes with more in-depth
study of the $\kappa=2$ case, as Ref.~\cite{seok} on this case has
studied only simpler specific cases. Since three position
parameters $z_l$'s can always be on a plane, we may take all
$z_l$'s to be points on the 1-2 plane.  Let us restrict our
attention to zeroes of Higgs appearing on $\mathbb{R}^3$
(corresponding to the choice $x^0 = 0$). Then all $X^0_l$'s and
other vectors related to the $x^0$-direction vanish, so that we
can express the determinant of $M$ by a simple form
\begin{eqnarray}\label{determinant}
\det M = | \vec{\cal F}  |^2 \left( | \vec{\cal F}  |^4  -
\sum_{ll'} (V_{ll'})_3 (W_{ll'})_3 | \vec{\cal F} |^2  +
2\sum_{ll'}(\vec{\cal F} \cdot \vec V_{ll'}) {\cal F}^3 (W_{ll'})_3
\right)~,
\end{eqnarray}
where the vector symbol means a vector in $\mathbb{R}^3$. Thus the
existence of zeroes requires that the right hand side of
(\ref{determinant}) should vanish. There are two possibilities, and
one is simply the zero total ``force" condition $\vec {\cal F} = 0
$. (In the expression (\ref{forces}), ${\cal F}^\mu_l$ has the same
form as 2-dimensional Coulomb force due to a source at $z_l$ with
charge $\rho_l^2$, and so we will call them ``forces" and $\cal F$ a
total force). This gives rise to zeroes which correspond generically
to two distinct points on the 1-2 plane, since there are three
sources for the force. The other possibility is that the quantity
inside the parentheses in (\ref{determinant}) may vanish: this case
is more complicated, and in fact contains the zeroes of the first
possibility. Thus we examined the second possibility to obtain
surfaces, on which the Higgs zeroes can lie, in Fig.~\ref{gp1} and
Fig.~\ref{gp2} with some choice of JNR parameters. The zero loci
found in Ref.~\cite{seok} are identified with the sections of the
surfaces in Figs.~\ref{gp1} and \ref{gp2} appearing on the 1-2 plane
($x^3=0$). Each of the sections are drawn in Fig.~\ref{gp1}-c) and
Fig.~\ref{gp2}-b), respectively. The isolated points at $x^3=0$ in
both surfaces correspond to the case with asymptotic values (with
fixed ${\rm v}=|\phi_0|$) $\vec\phi_0 =({\rm v}\cos\alpha,{\rm
v}\sin\alpha,0)$ for some $\alpha$, while the circle at $x^3=0$ in
Fig.~\ref{gp1}-c) and the closed loop at $x^3=0$ in
Fig.~\ref{gp2}-b) are the results with $\vec\phi_0 =(0,0,{\rm v})$.

Now let us concentrate on the most symmetric case shown in
Fig.~\ref{gp1}, and consider more complicated situation than that of
\cite{seok}, i.e., when we have nontrivial $(\phi_0^1,\phi_0^2)$
with nonzero value of $\phi_0^3$. Let us first focus on the section
of the zero surface appearing in the 1-3 plane ($x^2=0$) as shown in
Fig.~\ref{gp1}-d). By symmetry we here have ${\cal F}^2=0$ and
$\sum_{ll'} (V_{ll'})_2 (W_{ll'})_3=0$, so that the expression of
the Higgs field (and hence the matrix $M$ also) takes a simpler form
\begin{eqnarray}
\vec\phi(x) = \frac{1}{S\:\Pi(x)}
\begin{pmatrix}
    ({\cal F}^1)^2 - ({\cal F}^3)^2& 0 & 2{\cal F}^1 {\cal F}^3 + \sum_{ll'} (V_{ll'})_1 (W_{ll'})_3 \\
    0 & -(({\cal F}^1)^2 + ({\cal F}^3)^2) & 0 \\
    2{\cal F}^1 {\cal F}^3 & 0 & -(({\cal F}^1)^2 - ({\cal F}^3)^2) + \sum_{ll'} (V_{ll'})_3 (W_{ll'})_3
    \end{pmatrix}
            \begin{pmatrix}
              \phi_0^1 \\
              \phi_0^2 \\
              \phi_0^3 \\
            \end{pmatrix}~.\nn\\~
\end{eqnarray}
From the corresponding matrix $M$ one can easily see that
$\phi_0^2$ should vanish and $\phi_0^1$ is proportional to
$\phi_0^3$. If we start from $\vec\phi_0=(0,0,{\rm v})$ and
perform a rotation in the 1-3 plane (say, to $\vec\phi_0=({\rm
v}\sin\beta,0,{\rm v}\cos\beta)$) in group space, we can see the
zero locus lift away from the 1-2 plane. As we increase the angle
$\beta$ from zero to $\pi/2$, the part of the zero locus on the
1-3 plane moves along the curve depicted in Fig.~\ref{gp1}-d) from
the point A to the point C, via the point B. For this (most
symmetric) case, simultaneous rotations of the 1-2 and 3-0
coordinates by a same amount of angle are symmetries of the given
configuration. Thus the zero locus in this case (i.e., for a given
$\beta$) will be a circle which remains invariant under these
simultaneous rotations. This circle is a bit away from the 1-2
plane and also from the 3-0 plane\footnote{From the Higgs field
configuration (\ref{new form higgs}), we notice that the points in
the subsurface $\phi_0^2=0$ of the 3-dimensional hypersurface
$\det M=0$ can lie outside the subsurface $x^0=0$ of $\det M=0$.},
and as one can see from the figure, it shrinks to the point C as
one increases $\beta$ to $\pi/2$. We can also turn on a nonzero
$\phi_0^2$ at this stage. By symmetry, the zero locus will again
be a circle which remains invariant under the simultaneous
rotations of the 1-2 and 3-0 coordinates by a same amount of
angle. The intersecting point of this zero circle and the $x^0=0$
subsurface will be a generic point on the surface in
Fig.~\ref{gp1}-a).

\begin{figure}[t]
\centering\psfrag{p3}{$\vec \phi_0 = (0,0,\rm
v)$}\psfrag{A}{A}\psfrag{B}{B}\psfrag{C}{C} \psfrag{p13}{$\vec
\phi_0 = ({\rm v}\sin\beta,0,{\rm v}\cos\beta)$}\psfrag{p1}{$\vec
\phi_0 = (\rm v,0,0)$}
\includegraphics{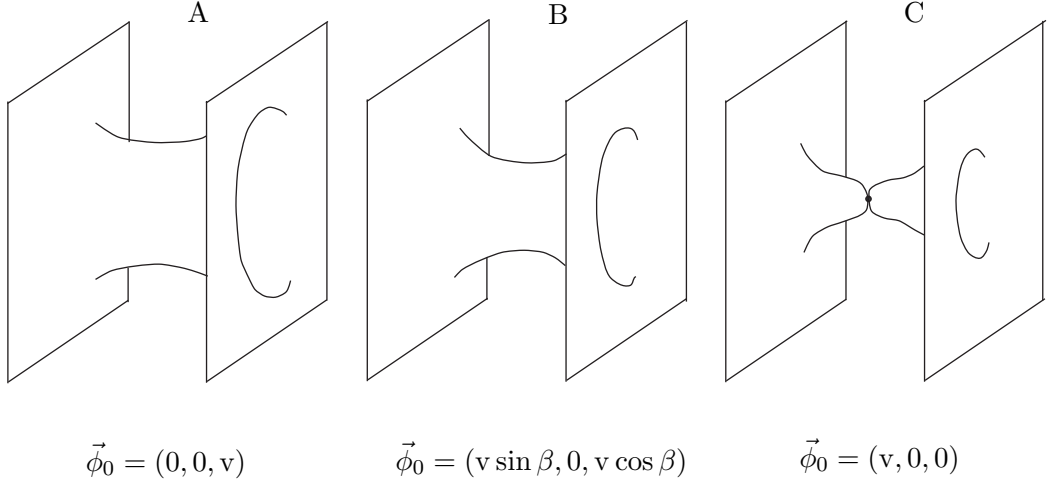}
\caption{The geometries of D4-brane-supertube configurations
corresponding to the cases described in Fig.~1. The cross sections
of the supertubes are always circluar in these cases. If we fix
the distance (i.e., set $|\phi_0|={\rm v}=$ const.) between two
D4-branes, the change from A to C becomes a rotation in the group
space. After this ninety degrees rotation, the supertube shrinks
to a point.}\label{gp11}
\end{figure}

As the eigenvalues of the asymptotic Higgs field can be interpreted
as the asymptotic positions (multiplied by 1/$i$) of two D4-branes,
we fix the asymptotic magnitude v but change the gauge orientation
of the Higgs field, which is equivalent to the changing of the gauge
orientation of instantons while fixing the Higgs orientation. The
deformation of the D4-branes indicated by the Higgs orientation
corresponding to the points A, B and C in Fig.~\ref{gp1} are
described by cartoon figures in Fig.~\ref{gp11}. The cross section
of the supertube collapses from a circle to a point as we change the
orientation of the Higgs field.

\begin{figure}[t]
\centering\includegraphics{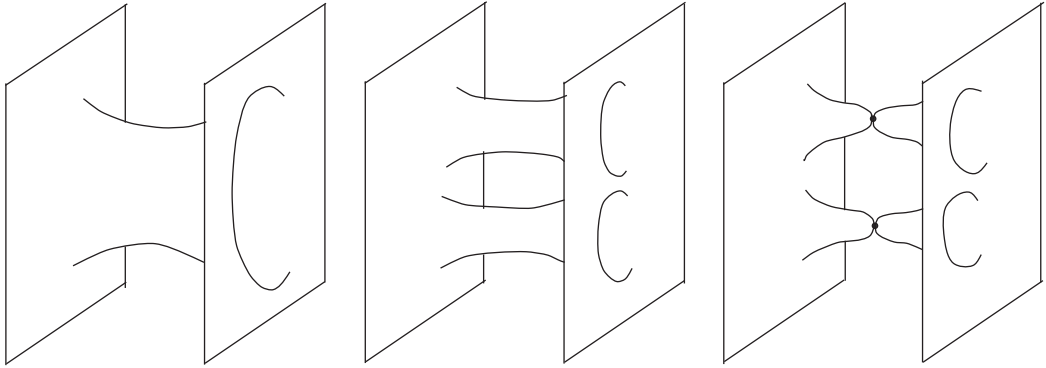}\caption{The geometries of
D-brane configurations corresponding to the situations described
in Fig.~2. In this figure, rotating $\phi_0$, a supertube splits
into two supertubes.}\label{gp22}
\end{figure}

For a more generic choice of JNR parameters, described in
Fig.~\ref{gp2}, we can again start from $\vec\phi_0=(0,0,{\rm v})$
and turn on nonzero $\phi_0^1,\phi_0^2$. As we noticed before, the
zero locus starts from a closed loop and ends up with two points
as we change the Higgs orientation from $(0,0,{\rm v})$ to $({\rm
v}\cos\alpha,{\rm v}\sin\alpha,0)$ ($\alpha$ is an arbitrary angle
here) in this case. The zero locus will get separated from the 1-2
plane and we can read the shape of the zero locus from
Fig.~\ref{gp2}. As we know the development of the Higgs zero locus
under rotations of the Higgs orientation, we can draw the cross
section of the supertube connecting D4-branes as in
Fig.~\ref{gp22}, which shows the break-up of a single cross
section to two cross sections.

\begin{figure}[t]
\centering\centering\psfrag{x1}{$x^1$}\psfrag{x2}{$x^2$}
\psfrag{x3}{$x^3$}\psfrag{x3=0}{$x^3=0$}
\includegraphics{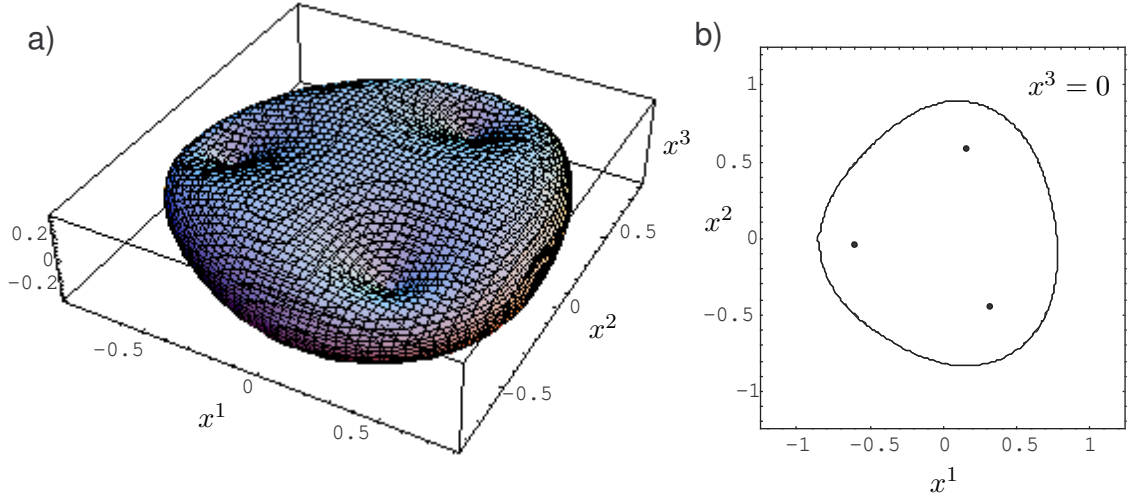}\caption{a) The surface $\det M=0$
at $x^0 =0$ with the instanton number $\kappa=3$,  $\rho_0 =0.8,
\rho_1 =1, \rho_2 =1.3, \rho_3 =1.2, z_0 =(-0.9,0), z_1 =(0,1),
z_2 =(1,0)$ and $z_3=(0,-1)$.  In general, four position
parameters $z_l$'s can always be located on a sphere $S^2$.
However, considering the facts that the JNR solutions are
conformally covariant and a plane is conformally related to $S^2$,
we can take the position parameters on a plane without an
excessive simplification. b) The section $x^3=0$ of the
surface.}\label{ggp3}
\end{figure}

\begin{figure}[t]
\centering\includegraphics{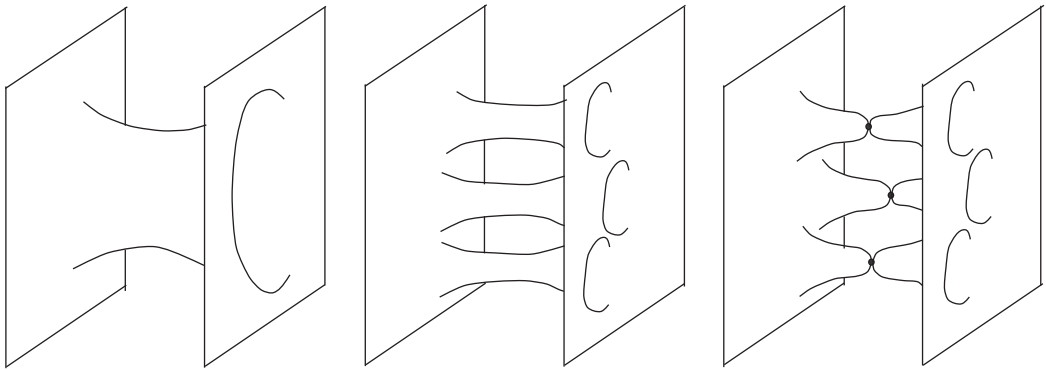}\caption{The D-brane
realizations of the situations in Fig.~5. We expect that there
happens splitting from one supertube to three
supertubes.}\label{ggp33}
\end{figure}

Next we discuss the zero locus in the Higgs solutions for the
$\kappa=3$ case. Here, although we can generally put four position
parameters $z_l$ only on $S^2$,  we put them on the 1-2 plane for
simplicity. In this case the determinant of the matrix $M$ is
again given by our formula (\ref{determinant}), from which we
obtain a surface of the Higgs zeroes as shown in Fig.~\ref{ggp3}.
As one can see from this figure, the zero locus of the Higgs field
will change from a closed curve to three isolated points on the
1-2 plane, as we change the orientation of the Higgs field from
$(0,0,{\rm v})$ to $({\rm v}\cos\alpha,{\rm v}\sin\alpha,0)$.
Related D4-brane-supertube configurations are given in
Fig.~\ref{ggp33} with three bridges between the two D4-branes.

From what we have found above, one may make a conjecture that the
zero locus of the Higgs field for the instanton number $\kappa$
with the $SU(2)$ gauge group can have at most $\kappa$
disconnected components. However, since we know neither the fully
explicit $\kappa$-instanton solutions nor the Higgs field
configurations representing dyonic instantons of the most general
type, it remains to be seen.

The JNR instanton solution has a residual local gauge symmetry
\cite{jnr} that we have already mentioned above. For any value of
$\kappa$, if all $\kappa + 1$ JNR position parameters are on a
circle or a line, there is a one-parameter family of JNR
parameters which are related by local gauge transformations.
Specifically, the $\kappa =2$ case has been explored in detail in
Ref.~\cite{atiyah}, and this one-parameter  family is called a
porism. In \cite{seok}, this porism structure was rederived using
the fact that zeroes of the Higgs field is gauge invariant. Now we
would like to understand the one-parameter gauge family of the
$\kappa =3$ JNR solutions in the same vein below.

For $\kappa = 3$, if we put all $z_l$'s on the 1-2 plane and take
$\phi_0^3 = 0$, then the second term of the Higgs field, i.e.,
$\phi_{\rm II}$ (see (\ref{phi2})) vanishes. So the condition for
Higgs zeroes becomes just the zero ``force" condition, $\vec{\cal
F} =0$. This zero ``force'' condition can be represented by a
complex equation
\begin{eqnarray}
\sum_{l=0}^{3} \frac{\rho_l^2}{ ({w - w_l})}=0~,
\end{eqnarray}
where $w \equiv x^1 + i x^2$ and  $w_l$'s denote the JNR position
parameters on a circle in the complex plane. We will write $w_l =
R e^{i \theta_l}$, $l=0,1,2$ here. We can rearrange this
condition, to write it by a complex cubic equation with three
complex parameters
\begin{eqnarray}\label{cubic}
w^3 -RC_1 w^2 + R^2 C_2 w -R^3 C_3=0~,
\end{eqnarray}
where
\begin{subeqnarray}
C_1 &=& \sum_{l=0}^3 (1-\lambda_l)e^{i\theta_l} ~,\\
C_2 &=& e^{i(\theta_0+\cdots + \theta_3)}\sum_{l=0}^{3}\sum_{k\neq
l}^{3} \lambda_l e^{-i(\theta_l + \theta_k)}~,\\ C_3 &=&
e^{i(\theta_0+\cdots + \theta_3)} \sum_{l=0}^{3} \lambda_l
e^{-i\theta_l}~.
\end{subeqnarray}
Here we have used notations $\lambda_l = \rho_l^2/S$
($S=\sum_{l=0}^3 \rho_l^2$), thus $\sum_{l=0}^3 \lambda_l=1$. Now
we would like to find a one-parameter family of the variation in
the JNR size and position parameters which can be identified with
a residual gauge transformation. As the Higgs field transforms
homogeneously under local gauge transformations, zeroes of the
Higgs field must be gauge invariant. Therefore, solutions of the
complex equation (\ref{cubic}) should not change if gauge
transformations are performed. Then the $C_i$'s ($i=1,2,3$) must
be invariant under gauge transformations. There are three complex
conditions for this invariance, i.e., $\delta C_{1,2,3} = 0$ and
seven parameters\footnote{These are four $\lambda_l$'s and four
$\theta_l$'s with a constraint $\sum_{l=0}^3 \lambda_l = 1$.}
which can be varied.  We are thus left with one-parameter family
of variation, related to the residual gauge transformation. If we
consider $ \delta \bar C_1 +e^{-i(\theta_0+\cdots
+\theta_3)}\delta C_3=0$, we get the condition
\be\label{condition} i\sum_{l} \left(\sum_k e^{-i\theta_k}
\lambda_k
 -e^{-i\theta_l}\right) \delta\theta_l \equiv i\sum_{l} A_l
\delta \theta_l = 0~.\ee
Considering the other equations $\delta C_2 = 0$ and $\delta C_3 =
0$, we can express $\delta \lambda_l$'s as linear combinations of
$\delta \theta_l$. Due to the constraint $\sum_l \lambda_l=1$,
$\sum_l \delta \lambda_l$ should vanish, so the equations which we
have to solve are one complex equation (\ref{condition}) and
\begin{eqnarray}\label{lambda}
\sum_l \delta \lambda_l \equiv \frac{\sum_l B_l \delta
\theta_l}{\cos(\frac{\theta_0 +\theta_1 -\theta_2 -\theta_3}{2}) +
\cos(\frac{\theta_0 +\theta_2 -\theta_1 -\theta_3}{2} ) +
\cos(\frac{\theta_0 +\theta_3 -\theta_1 -\theta_2}{2} )   }  =
0~~,
\end{eqnarray}
where $B_l$ is given by
\begin{eqnarray}\nn
B_l = \frac{i}{2}e^{\frac{i}{2} (\theta_0 + \ldots
+\theta_3)}&\Bigg[& \sum_k\lambda_k e^{-2i\theta_k} -2\lambda_l
e^{-2i\theta_l} + e^{-i \theta_l}\big(\sum_k \lambda_k e^{-i
\theta_k} \big) \\&& -\big(\sum_k
e^{-i\theta_k}\big)\big(\sum_{l'} \lambda_{l'} e^{-i \theta_{l'}}
- \lambda_l
 e^{-i \theta_l}\big) ~\Bigg]\\\nn &+& ~~\text{complex conjugate}.
\end{eqnarray}
The equations (\ref{condition}) and (\ref{lambda}) are solved by
choosing the parametrization
\begin{eqnarray}
\delta \theta_l = i \sum_{l_1,l_2,l_3 =0}^3 \varepsilon^{l\: l_1
l_2 l_3} A_{l_1} \bar A_{l_2} B_{l_3}~ \delta \tau ~.
\end{eqnarray}
Thus we have obtained an one-parameter family of variations. We
believe that this one-parameter family of variations corresponds
to the residual gauge symmetry of JNR instantons; but, to be
definite, some further check will be required.

%
%
This consideration is actually valid even for the general $\kappa$
case. In the general case we have a $\kappa$-th order complex
equation with $\kappa$ complex parameters, and there are also $2
\kappa + 1$ JNR position and size parameters on a circle;
therefore, we are left with an one-parameter family of gauge
transformations.


\section{Conclusion and Discussions}

In this work we have studied BPS dyonic instantons in the Coulomb
phase, and presented  a general formalism to find the Higgs field
satisfying the covariant Laplace equation in the general ADHM
framework for the instantons. Especially, we found the explicit
expression for the Higgs field solution in the Jackiw-Nohl-Rebbi
three- or four-instanton background. In addition we explored in
detail the zero locus of the Higgs field for two (and three,
partially) instanton case and studied some aspect of the residual
gauge freedom for JNR three-instanton solutions.

Our analysis shows that the Higgs solution in the instanton
background has a very rich structure. While our detailed study was
restricted to the case with the Jackiw-Nohl-Rebbi three- of
four-instanton backgrounds, the general three-instanton solution
has been found in Refs.~\cite{cws,korepin} and the structure of
related Higgs solutions need to be analyzed also.

Our dyonic instantons have  very large degeneracy even when the
instanton number and the electric charge are fixed. The moduli
space dynamics of dyonic instantons is a phase space dynamics in
the sense that the first order in  time derivative term dominates.
(This can be seen easily from the moduli-space dynamics of
instantons). As explained in Ref.~\cite{tong}, the moduli space
dynamics of the instanton is also corrected by a potential term
given by the Killing vector related to the symmetry breaking. The
electric charge interaction and the potential are thus balanced
and for the resulting BPS configuration the relative motion moduli
space dynamics becomes first order in time. The supersymmetric
generalization and detailed exploration of this dynamics needs a
further consideration.

In a similar way to supertubes, dyonic instantons also carry
nonzero angular momentum in 4-dimensions. One may split the
angular momentum value into the self-dual and anti-self-dual
parts. The detailed evaluation of the angular momentum in terms of
the ADHM data remains to be done. There is a close relation
between the shape of the zero locus and the magnitude of the
angular momentum. For a given instanton number and electric
charge, one expect that there exists an upper bound on the angular
momentum. It can easily be estimated by the supertube analysis
done in Ref.~\cite{leeyee}. In the large instanton limit and a
circular magnetic monopole string case, one can approximate the
magnetic monopole string as a straight string locally. For a given
asymptotic Higgs expectation value v, the tension of the string
and the momentum density is fixed as $4\pi {\rm v}/e^2$. There is
one parameter $h$ which allows to write the instanton energy
density and charge density as $4\pi h/e^2$ and  $4\pi {\rm v}/(e^2
h)$, respectively, so that their product is independent of $h$.
For a circle-shaped monopole string of radius $R$ lying on the 1-2
plane, the total instanton number would be $\kappa =Rh$ so that
its energy becomes ${\cal E}=8\pi^2Rh/e^2$, and the total electric
charge would be $Q_e=8\pi^2 {\rm v}R/(e^2h)$. The estimated value
of the maximal angular momentum would then be
\be J_{12} = \frac{8\pi^2}{e^2}{\rm v}R^2= \kappa Q_e~, \ee
which is independent of $R$ (as $\kappa$ and $Q_e$ are kept fixed)
and so of the density. It would be nice to verify that the maximal
value of the angular momentum is indeed of this form. Related to
this, the angular momentum in the $\kappa=2$ case was studied by
the authors of Refs.~\cite{hash,sungjay} in a field theory and in
a supergravity model respectively.

\acknowledgments{We would like to thank Seok Kim and Sungjay Lee
for useful discussions. The work of M.-Y.C. and C.L. was supported
by the Korea Science Foundation ABRL program (R
14-2003-012-01002-0). The work is also supported in part by the
KOSEF SRC Program through CQUeST at Sogang University (K.-M.L.),
KRF Grant No. KRF-2005-070-C00030 (K.-M.L) and the National
Scholar Program of KRF (K.K.K., K.-M.L.).}

\appendix

\section{Direct Verification of our Higgs Solution}

In the main text our Higgs solution (\ref{main2}) in the ADHM
background was obtained from the asymptotic limit of the related
scalar propagator. We shall here provide a direct check on this
result by showing that the expression (\ref{main2}) indeed solves
the covariant Laplace equation (\ref{cL2}). For this purpose it is
convenient to write the result (\ref{main2}) in the form
\be\label{w1} \phi(x)=v^\dag (x)Wv(x)~,\ee
introducing a $x$-independent $(\kappa+1)\times(\kappa+1)$ matrix
of quaternions
\be\label{w2} W=V\phi_0 V^\dag -2C{\cal A}C^\dag~.\ee
Also useful is the well-known result that, as we represent the ADHM
constraint by $\Delta^\dag(x)\Delta(x)=f^{-1}(x) e_0$ ($f(x)$ is a
real, invertible $\kappa\times\kappa$ matrix), we have
\cite{adhm,cws,cori}
\be\label{proj} 1-v^\dag(x)v(x)=\Delta (x) f(x) \Delta^\dag
(x)~.\ee

From (\ref{w1}) and (\ref{w2}), it is not difficult to derive
\ba D_\mu \phi &=& \p_\mu v^\dag (1-vv^\dag)Wv +v^\dag W
(1-vv^\dag)\p_\mu v\nn\\&=& -v^\dag (\p_\mu \Delta)f\Delta^\dag
Wv-v^\dag W\Delta f(\p_\mu \Delta^\dag)v~,\ea
where we used the relation (\ref{proj}) as well as the equations
in (\ref{constraint}). Then taking the covariant derivative once
more with the thus-obtained expression, we find, after somewhat
lengthy algebra, the following expression:
\be\label{cL3} D_\mu D_\mu \phi = -4v^\dag \{CfC^\dag,W\}v
+4v^\dag Cf\cdot {\rm tr} \Delta^\dag W\Delta \cdot fC^\dag v~.\ee
To obtain this form, we made use of the fact that $e_\mu q \bar
e_\mu = 2{\rm tr}\: q$ for any quaternion $q$.

We may now insert the expression (\ref{w2}) for $W$ into the right
hand side of (\ref{cL3}). Then, due to the fact that $C^\dag V=
V^\dag C=0$ (see (\ref{asym4})), the first piece in (\ref{cL3})
can be written as
\be\label{1st} -4v^\dag \{CfC^\dag,W\}v = 8v^\dag C(fC^\dag C{\cal
A}+{\cal A}C^\dag Cf)C^\dag v~.\ee
Also, if the expression (\ref{w2}) is used in the ${\rm tr}$-term
from the second piece of (\ref{cL3}), it can be reduced to
\ba {\rm tr} \Delta^\dag W\Delta &=& {\rm tr} B^\dag V\phi_0
V^\dag B-2C^\dag C{\cal A}f^{-1} -2f^{-1}{\cal A}C^\dag C\nn\\&~&
-{\rm tr} \left\{B^\dag [C{\cal A},C^\dag]B+[B^\dag,C^\dag C{\cal
A}]B +B^\dag [C^\dag, {\cal A}C^\dag]B +B^\dag[{\cal A}C^\dag C,
B]\right\}~.\nn\\\label{tr2}~\ea
To obtain this result, we made use of the observation that all
quadratic and linear terms in $x$ from the expression
\be {\rm tr} \left\{ \Delta^\dag [C{\cal A},C^\dag]\Delta
+[\Delta^\dag, C^\dag C{\cal A}]\Delta +\Delta^\dag [C, {\cal
A}C^\dag]\Delta +\Delta^\dag [{\cal A}C^\dag C,\Delta]\right\} \ee
cancel, to leave only the $x$-independent contribution equal to
the last ${\rm tr}$-term in (\ref{tr2}). As we use the results
(\ref{1st}) and (\ref{tr2}) in (\ref{cL3}), we are then left with
the expression
\ba D_\mu D_\mu \phi &=& 4v^\dag Cf\cdot \left[{\rm tr} B^\dag
V\phi_0 V^\dag B-{\rm tr} \left\{ B^\dag [C{\cal
A},C^\dag]B+[B^\dag,C^\dag C{\cal A}]B
\right.\right.\nn\\\label{cL4}&~& \left.\left. +B^\dag [C,{\cal
A}C^\dag]B +B^\dag[{\cal A}C^\dag C, B]\right\}\right]\cdot
fC^\dag v~.\ea
If we here define a quantity $R$, an antisymmetric
$\kappa\times\kappa$ matrix, by
\be R={\rm tr} \left\{2B^\dag C{\cal A}C^\dag B -C^\dag C{\cal
A}B^\dag B - B^\dag B {\cal A} C^\dag C\right\} \ee
and a $\kappa$-column vector $\tilde v = 2fC^\dag v$, (\ref{cL4})
can be further simplified to
\be\label{cL5} D_\mu D_\mu \phi = {\tilde v}^\dag\cdot\left[{\rm
tr} B^\dag V\phi_0 V^\dag B -R\right]\cdot \tilde v  ~.\ee

Based on (\ref{cL5}), we conclude that the Higgs configuration
(\ref{w1}) corresponds to the solution of (\ref{cL2}) only if the
matrix ${\cal A}=({\cal A}_{ij})$ satisfies the linear
inhomogeneous equations
\ba &~& {\rm tr} (B^\dag V)_m \phi_0 (V^\dag B)_n~=~R_{mn}
\nn\\&~&~~~ =~ \frac12{\rm tr}\left\{2B^\dag C{\cal A}C^\dag B
-C^\dag C{\cal A}B^\dag B - B^\dag B {\cal A} C^\dag
C\right\}_{mn} -(m\leftrightarrow n)\\&~&~~~ = \left[\frac12{\rm
tr} \left\{ 2(C^\dag B)_{mr} (B^\dag C)_{sn}-(C^\dag C)_{mr}
(B^\dag B)_{sn} - (B^\dag B)_{mr} (C^\dag
C)_{sn}\right\}-(m\leftrightarrow n)\right]{\cal A}_{rs}~.\nn \ea
These coincide with the equations we found for ${\cal A}$ in the
main text, (\ref{A}). This completes the verification.

\section{Computation of Electric Charge}

We shall here present the derivation of our expression
(\ref{adhmchg}) for the electric charge of dyonic instantons.
General $SU(2)$ dyonic instantons are described by the ADHM
instantons (see (\ref{gauge}) -- (\ref{Delta})) and the
corresponding Higgs field, written conveniently in our form
(\ref{w1}) with matrix $W$ (see (\ref{w2})). Then, thanks to the
BPS equations (\ref{BPS}), we can express the electric field
$E_\mu$ as
\ba E_\mu
&=& D_\mu\phi ~=~ \p_\mu \phi + [v^\dag\p_\mu v ,\phi]\nn\\
&=& (\p_\mu v^\dag) Wv +v^\dag W \p_\mu v-( \p_\mu v^\dag ) v
v^\dag Wv-v^\dag Wv v^\dag \p_\mu v~.\ea
Using (\ref{proj}), this can be rewritten in the form
\ba E_\mu &=& -v^\dag (\p_\mu \Delta)f\Delta^\dag Wv -v^\dag
W\Delta f(\p_\mu \Delta^\dag)v \nn\\&=& v^\dag(Ce_\mu)f(B^\dag
-\bar x C^\dag)Wv+v^\dag W(B-Cx)f(\bar e_\mu C^\dag)v~,\ea
where we used also the relation $(\p_\mu v^\dag)\Delta = -v^\dag
\p_\mu \Delta$ and $\Delta^\dag \p_\mu v=-(\p_\mu \Delta^\dag)v$
(following from (\ref{constraint})). Inserting (\ref{w2}) for $W$
and using $C^\dag V = V^\dag C = 0$, it is possible to recast the
above expression in the form
\ba E_\mu &=& v^\dag C e_\mu f B^\dag V\phi_0 V^\dag v +v^\dag
V\phi_0 V^\dag Bf\bar e_\mu C^\dag v \nn\\&~&-2v^\dag Ce_\mu
f(B^\dag C-\bar x C^\dag C){\cal A}C^\dag v -2v^\dag C{\cal
A}(C^\dag B-C^\dag Cx)f\bar e_\mu C^\dag v~,\ea
which is rather complicated, but contains no derivative.

The electric charge (\ref{charge}) is given by the surface
integral
\be\label{chgint} Q_e = -\oint_{S^3_\infty} {\rm d}S_\mu ~
\frac1{\rm v}{\rm tr}(\phi E_\mu)=-\lim_{|x|\rightarrow\infty}
\int {\rm d}\Omega(\hat x)~|x|^3 \hat x_\mu ~ \frac1{\rm v}{\rm
tr}(\phi E_\mu)~.\ee
Taking the asymptotic behaviors (\ref{asym2})--(\ref{asym3}) and
that of $f(x)$
\be f(x) = \frac1{|x|^2}f_0 +{\cal O}\left(\frac1{|x|^3}\right)~,
~~~(f_0 \equiv (C^\dag C)^{-1})\ee
(as follows from studying $\Delta^\dag \Delta= (B^\dag -\bar x
C^\dag)(B-Cx)=f^{-1}$) into account, we can get the asymptotic
behavior of $E_\mu$
\ba E_\mu &\sim& \frac1{|x|^3} \left[\bar g V^\dag B\hat{\bar
x}e_\mu f_0 B^\dag V\phi_0 g +\bar g \phi_0 V^\dag Bf_0 \bar e_\mu
\hat x B^\dag Vg\right]\nn \\&~& + \frac1{|x|^4}\left[2\bar g
V^\dag B \hat{\bar x} e_\mu f_0 (\bar x C^\dag C-B^\dag C) {\cal
A} \hat x B^\dag Vg +2\bar g V^\dag B\hat{\bar x}{\cal A}(C^\dag C
x-C^\dag B) f_0 \bar e_\mu \hat x B^\dag Vg \right]~.\nn\\~ \ea
Based on this result and (\ref{hvev}), one finds the following
expression for the integrand of (\ref{chgint}):
\ba |x|^3 \hat x_\mu {\rm tr}(\phi E_\mu) &=& |x|^3 \hat x_\mu
{\rm tr} \left\{(\bar g\phi_0 g)\cdot\frac1{|x|^3} \left[\bar g
V^\dag B\hat{\bar x}e_\mu f_0 B^\dag V\phi_0 g +\bar g \phi_0
V^\dag Bf_0 \bar e_\mu \hat x B^\dag Vg\right.\right.\nn\\&~&
\left.\left. +2\bar g V^\dag B \hat{\bar x} e_\mu f_0 \hat{\bar x}
C^\dag C {\cal A} \hat x B^\dag Vg +2\bar g V^\dag B\hat{\bar
x}{\cal A}C^\dag C\hat x f_0 \bar e_\mu \hat x B^\dag Vg
\right]+{\cal
O}\left(\frac1{|x|^4}\right)\right\}~.\nn\\\label{B7}~\ea
Using the cyclic property of the trace, $\bar g g=1$ and
$\hat{\bar x}\hat x =1$ ($g$ and $\hat x \equiv \hat x_\mu e_\mu$
are unit quaternions), a very simple expression
\be \lim_{|x|\rightarrow \infty} |x|^3\hat x_\mu {\rm tr} (\phi
E_\mu) = {\rm tr} \left\{2\phi_0^2 V^\dag B (C^\dag C)^{-1} B^\dag
V  +4\phi_0 V^\dag B {\cal A} B^\dag V \right\}~\ee
can be obtained from (\ref{B7}). From this asymptotic form one can
easily deduce that the electric charge is given by the form
(\ref{adhmchg}).

\end{document}